\documentclass[aps,pre,twocolumn]{revtex4-1}

\usepackage{bm}
\usepackage[dvipdfmx]{graphicx}
\usepackage{here}
\usepackage{amsmath}
\usepackage{mathrsfs}
\usepackage{amssymb}

\usepackage{color}

\begin{document}

\title{Surface-tension-driven coarsening in mass-conserved reaction-diffusion systems}
\author{Michio Tateno}
\email{c-tateno@g.ecc.u-tokyo.ac.jp} 
\author{Shuji Ishihara}
\affiliation{Graduate School of Arts and Sciences, The University of Tokyo, Komaba 3-8-1, Meguro-ku, Tokyo 153-8902, Japan}
\affiliation{Universal Biology Institute, The University of Tokyo, Komaba 3-8-1, Meguro-ku, Tokyo 153-8902, Japan}
\date{\today}

\begin{abstract}
{\bf
Mass conservation in chemical species appears in a broad class of reaction-diffusion systems (RDs) and is known to bring about coarsening of the pattern in chemical concentration. 
Recent theoretical studies on RDs with mass conservation (MCRDs) reported that the interfacial curvature between two states contributes to the coarsening process, reminiscent of phase separation phenomena. 
However, since MCRDs do not presuppose a variational principle, it is largely unknown whether description of surface tension is operative or not. 
In this study, we numerically and theoretically explore the coarsening process of patterns in MCRDs 
in two and three dimensions. 
We identify the parameter regions where the homogeneous steady state becomes stable, unstable, 
and metastable. In the unstable region, pattern formation is triggered by usual Turing instability, whereas in the metastable region, nucleation-growth-type pattern formation is observed. 
In the later stage, spherical droplet patterns are observed in both regions, where they obey a relation similar to the Young-Laplace law and coarsen following the evaporation-condensation mechanism. 
These results demonstrate that in the presence of a conserved variable, a physical quantity
similar to surface tension is relevant to MCRDs, which provides new insight into molecular self-assembly driven by chemical reactions.
}
\end{abstract}

\maketitle

\noindent
\section{Introduction} 
Reaction-diffusion systems (RDs) are one of the most generic mathematical frameworks 
that give rise to spatio-temporal patterns and have been applied to various phenomena, 
ranging from physics, chemistry, biology, geology, to ecology \cite{cross_pattern_1993, bressloff_what_2002, hoyle_pattern_2006, meron_pattern_2016}. 
The Turing mechanism explains how a homogenous initial state is destabilized to develop a spatial structure; infinitesimally small fluctuations with a particular wave number $q_{\rm m}$ selectively grows exponentially. Notably, despite this analysis being valid only for every early linear time regime, 
the length scale obtained from this analysis $\ell_{\rm m}(\equiv 2\pi/q_{\rm m})$ usually provides a good estimation of the characteristic length scale of final steady-state patterns, such as stripes or hexagonal dots in two-dimensional (2D) systems.

Mass-conserved reaction-diffusion systems (MCRDs) do not follow the above trend.
MCRDs were originally developed to model molecular localization of membrane-bounded proteins
inside cells, specifically Rho-family GTPases, which regulate cell polarity
~\cite{otsuji_mass_2007, ishihara_transient_2007, goryachev_dynamics_2008, mori_wave-pinning_2008, rubinstein2012weakly, holmes_analysis_2016, chiou_principles_2018, jacobs_small_2019, brauns2018phase, brauns2020wavelength} (see also a recent review article Ref.~\cite{goryachev2020compete}). 
In such systems, it was found that the characteristic length scale of the patterns easily grows beyond $\ell_{\rm m}$ after a linear time regime and often reaches a length scale comparable to the system size $L$.
In the context of biomolecular self-assembly, this feature indicates that 
proteins accumulate at one site, not spreading over multiple sites, 
and is crucial to determining the unique directionality of cells.
It was also discussed that the resulting scalability of the pattern against the system size has biological significance in morphogenesis~\cite{othmer_scale-invariance_1980, hunding_size_1988, ishihara_turing_2006, ben-zvi_scaling_2010, wartlick_dynamics_2011, umulis_mechanisms_2013}. 

\begin{figure*}[t!]
\centering
\includegraphics[width=16.0cm]{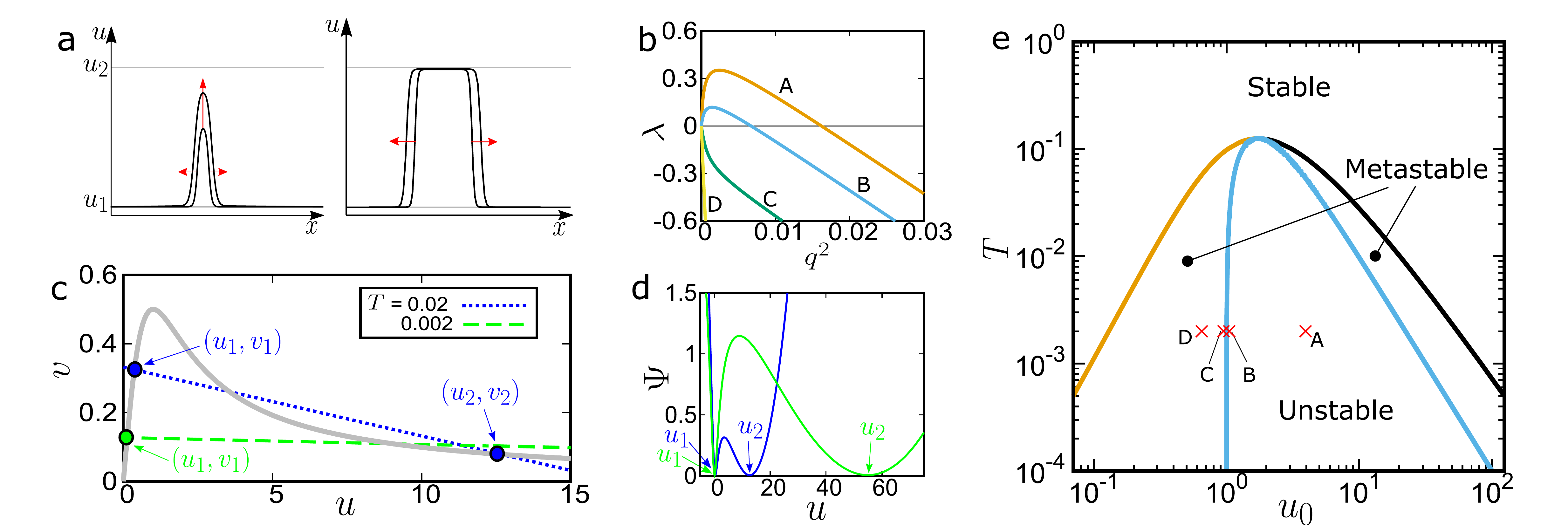}
\caption{
{\bf Steady state analysis of MCRD systems.}
{\bf a}. Two typical density profiles in MCRDs. $u_1$ and $u_2$ denote $u$ at the bistable points. Depending on whether $u$ in the domain is saturated to the stable points or not, the density profile of $u$ show ``mesa-like'' (wave-pinning type, left panel) 
or ``peak-like'' (Turing type, right panel) behavior. 
{\bf b}. Dispersion relation in the homogeneous state. 
The largest growth rate $\lambda$ is shown for various parameter values indicated by the labels corresponding to those in panel {\bf e}. For large scale ($q \rightarrow 0$), 
$\lambda$ follows $\lambda = -(D_v f_u - D_u f_v)/(f_u - f_v)q^2 + \mathscr{O}(q^4)$, and the sign of the prefactor of $q^2$ determines the stability of the homogeneous state. 
{\bf c}. Mapping of the steady stable points on the $u$-$v$ space for two different parameter values of $T$. The gray curve is the nullcline for the reaction term $f$. The blue and green lines represent $p_{\infty} = D_u u + D_v v$ for $T=0.02$ and $0.002$, respectively. The blue and green filled circles represent the stable fixed points. Note that $f$ is independent of the choice of the model parameters in our normalization (see Appendix~B). 
{\bf d}. Function profiles of $\Psi$ for $T=0.02$ (blue) and $0.002$ (green) at $p=p_\infty$. 
{\bf e}. Phase diagram. The blue curve represents the border of the unstable 
region where the Turing mechanism is operative. 
The black and gray curves represent the set of stable points, $u_1$ and $u_2$, respectively. 
We denote the region outside of the two curves as ``stable'' and regions that are neither stable nor unstable as ``metastable''. 
}
\label{fig:fig0}
\end{figure*}

Pattern formation of MCRDs in one dimension has been intensively investigated in previous research.
Two types of MCRDs were mainly studied: 
in the first case, referred to as the ``Turing type'' \cite{chiou_principles_2018}, multiple peak-like domains, illustrated in the right panel of Fig.~\ref{fig:fig0}a, appear from a homogeneous initial state via the Turing mechanism
\cite{otsuji_mass_2007, ishihara_transient_2007}; 
the second case is the ``wave-pinning type''~\cite{mori_wave-pinning_2008}, where 
the system is bistable with high and low chemical density states, and
by exposing a finite (not infinitesimally small) perturbation to
the homogeneous state, mesa-like concentration profiles emerge as illustrated in the left panel of Fig.~\ref{fig:fig0}a.
Both systems exhibit coarsening, i.e., smaller peaks and mesas shrink and disappear while 
larger ones grow. 
For the Turing type, both the height and width of the peak increase, whereas for the wave-pinning type, only the width of the mesa grows while keeping the height (see Fig.~\ref{fig:fig0}a).
For a long time scale, the system reaches a single isolated domain in the Turing type. On the other side, for the wave-pinning type, the decrease in the number of domains significantly slows down at some point, resulting in the apparent coexistence of multiple domains.  

Despite the difference between the Turing and wave-pinning types mentioned above, recent studies revealed that these differences would be rather superficial~\cite{chiou_principles_2018, jacobs_small_2019, brauns2020wavelength}. 
Chiou~et~al.~\cite{chiou_principles_2018} found that the peak-like profile in Turing type MCRDs is owing to the unsaturation of the peak height.
That is, the peak-like pattern is transient and
it will eventually become mesa-like for a sufficiently large system size and long simulation time. 
Their study also discussed that the difference in the peak heights causes a difference in 
``recruitment power'', which enhances the flux of the chemical component from lower 
peaks to higher ones (see $\nabla p$ introduced in Sec.~II). This drives faster coarsening of peaks seen in the Turing type than the wave-pinning type. 
Furthermore, Turing instability can occur in the wave-pinning type MCRDs by choosing appropriate parameter values~\cite{holmes_analysis_2016, chiou_principles_2018, jacobs_small_2019}.
Taken together, it is understood that there is quantitative, but no essential difference
 between long term dynamics of two cases.
We note that the coexistence of multiple peaks can also be induced by a slight violation of mass 
conservation \cite{jacobs_small_2019, brauns2020wavelength} or the introduction of negative feedback~\cite{jacobs_small_2019}.

However, since earlier studies have conducted intensive analyses only in 1D systems, there remains a fundamental untested component in understanding the dynamics of MCRDs. For higher dimensions, the interfacial morphology of patterns may be crucial to the pattern formation dynamics of MCRDs. 
Indeed, Refs.~\cite{chiou_principles_2018, jacobs_small_2019, brauns2020wavelength} pointed out that 
the curvature of the interface affects the coarsening of the patterns in the late stage. 
This observation raises a possible coarsening mechanism governed by ``surface tension'' 
in MCRDs, similar to the phase separation phenomena~\cite{onuki2002phase}. 
However, in MCRDs the chemical reaction violates the detailed balance condition in general, and the existence of the variational functional, which is prerequisite for the theory of phase separation, is not obvious. 
Thus, it is unclear whether the physical description of the surface tension could be rationalized. 

Motivated by the question raised above, this study explores 
the influence of curved interface on pattern formation dynamics in MCRDs.
We perform extensive numerical simulations using multiple GPUs for both 2D and 3D systems, and 
the obtained dynamics are analyzed using methods developed for phase separation phenomena, 
together with technique of dynamical systems theory. 
By investigating the similarities and differences between MCRDs and phase separation phenomena, we explore whether mathematical structures such as energy variational principles are also inherent in MCRDs. 

\section{System description}
We consider a system composed of two chemical species, $U$ and $V$, and denote their
concentration fields as $u(t, {\bm r})$ and $v(t, {\bm r})$, respectively. 
These species are interconvertible by chemical reactions
$U \rightarrow V$ and $V \rightarrow U$, the rates of which are $k_u u$ and $k_v v$, respectively. 
Note that these two reactions do not have to be in reverse relationship
and the system may not satisfy detailed balance.
Both the transition rates $k_u$ and $k_v$ can be dependent on the density states $u$ and $v$, 
as a result of positive feedback regulation in the chemical reaction network.
Moreover, both $U$ and $V$ migrate over space by thermal diffusion. 
We model the time-evolution equations for $u$ and $v$ as the following reaction-diffusion equations:
\begin{eqnarray}
\frac{\partial u}{\partial t} &= D_u \nabla^2 u - f(u, v), \label{eq:RD1_1} \\
\frac{\partial v}{\partial t} &= D_v \nabla^2 v + f(u, v). \label{eq:RD1_2}
\end{eqnarray} 
Here, $D_u$ and $D_v$ are the diffusion coefficients of $U$ and $V$, respectively, and we assume $D_u < D_v$. $f$ is the reaction term and written as $f = k_u u - k_v v$.  
By adding the above two equations, we obtain
\begin{equation}
\frac{\partial s}{\partial t} = \nabla^2 p, 
\label{eq:conseved}
\end{equation}
where $s = u + v$ and $p = D_u u + D_v v$. Eq.~\ref{eq:conseved} indicates that $s$ is a conserved variable without material sources. 

The relevance of adopting MCRDs to describe cellular processes 
can be rationalized as follows: 
(i)~Many subcellular processes, such as cell polarity formation, are sufficiently rapid
($\sim$1 min) compared with the production and degradation rates of the molecules
($>$10 min). 
Such processes are usually conducted only through the activation and deactivation of molecules, such as 
phosphorylation and dephosphorylation. 
The ratio of molecules between ``on'' and ``off'' states changes; however, the total mass of the molecule is conserved throughout the processes. 
(ii)~Molecular interaction contains positive feedback loops that regulate the switching rate between molecular states, e.g., the acceleration of phosphorylation is found in subcellular processes. 
(iii)~Molecular states are often associated with an affinity to the cellular membrane; thus, molecules in different states are likely to attach to or detach from the membrane. As a result, molecular diffusion coefficients are significantly different between the molecular states, being small on the cell membrane but large in the cytosol. 
As these processes are basic, many previous models, explicitly or implicitly, contain chemical components that are conserved through dynamics. In Ref.~\cite{otsuji_mass_2007}, it was shown that in a model with these conditions, a uniform distribution of molecules can be destabilized in a Turing-like manner, leading to an accumulation of molecules at one site. 

To examine the generality of the results presented below, we employ two models, following earlier works on MCRDs, both of which violate the detailed balance condition. 
Model~I employs $k_u = ca/(b+u^2)$ and $k_v =c$ \cite{otsuji_mass_2007, ishihara_transient_2007}.
For Model~II, $k_u = \alpha$ and $k_v =k_0 + \beta u^2/(K^2+u^2))$ are employed \cite{mori_wave-pinning_2008}. 
The former and latter models correspond to Turing type and wave-pinning type, respectively, 
by the criteria mentioned in the Introduction. 
Because these models share many aspects in their dynamical trends, we primarily describe the results of Model I in the main text, and make comments when there are remarkable differences in the results between the two models. The results from Model II are provided in Appendix~D. 

In the results presented below, the statistical data are based on simulations performed on $4096^2$ and $1024^3$ grids for 2D and 3D systems, and have normalized units, described in Appendix~B. 
After normalization the governing equations Eq.~\ref{eq:RD1_1} and \ref{eq:RD1_2} have two free parameters:
one is associated with the reaction term and is defined as $R=a/b$ and $\alpha/\beta$ for Model I and II, respectively; the other is $D=D_v/D_u$, associated with the diffusion constants.
As we will show below, the stability of the homogeneous state is determined by the product of these parameters $T^{-1} \equiv RD$, and the initial chemical concentrations that determine the conserved quantity in the system.

\section{Results}
\subsection{Phase diagram of MCRDs}
Before investigating pattern formation dynamics in MCRDs, 
we begin with identifying their parameter dependencies by analyzing the initial homogeneous and eventual bistable solutions.

We first consider the stability of the homogeneous steady solution $(u_0, v_0)$. $u_0$ and $v_0$ are determined by the conditions $f(u_0, v_0)=0$ and  $s_0=u_0+v_0$, where $s_0 \equiv \int s d{\bm r}/V$ is the mean molecular concentration conserved through the dynamics (see Eq.~\ref{eq:conseved}), and $V$ is the volume of the system.
We suppose that the fixed point $(u_0, v_0)$ is linearly stable when the diffusion terms are absent in Eqs.~\ref{eq:RD1_1} and~\ref{eq:RD1_2} (otherwise the system runs to infinity). This is conditioned 
by $f_u - f_v >0$, where $f_u$ and $f_v$ represent $\partial f/\partial u$ and $\partial f/\partial v$,
respectively. 
In the presence of the diffusion terms, 
the homogeneous state becomes linearly unstable against an infinitesimally small 
perturbation with wave number $q$ between $0<q^2<(D_u f_v - D_v f_u)/D_u D_v$ when 
$(D_v f_u - D_u f_v)/(f_u - f_v)<0$ is satisfied.
This condition corresponds to the Turing instability.
Since the denominator is positive as mentioned above, Turing instability condition is simply written as
\begin{equation}
\label{eq:TuringCond}
	D_v f_u - D_u f_v < 0 ~.
\end{equation}
For this condition to hold for Model I, $T < T_{\rm c} \equiv 1/8$ is necessary (see Appendix~B).
The dispersion relation obtained in the analysis is shown
in Fig.~\ref{fig:fig0}b, where $\lambda(q)$ represents the growth rate
of the perturbation with wave number $q$. 
Note that, owing to mass conservation, $\lambda(q)$ always starts from zero in MCRDs unlike typical RDs.

Next, we explore the dynamic behavior in the long time regime.
The most significant feature in MCRDs is the existence of a conserved variable $s$, and the transport of $s$ is driven by the spatial gradient of a scalar variable $p$ as indicated by Eq.~\ref{eq:conseved}. 
At the steady state, $p$ satisfies $\nabla^2 p=0$. This implies that $p$ becomes a constant value $p_{\infty}$ in the absence of material source.
Owing to the constant nature of $p$, 
$(u, v)$ is constrained on a line $p_{\rm \infty} = D_u u + D_v v$ which has three intersection points with the nullcline $f=0$ (see Fig.~\ref{fig:fig0}c). 
Two of them, indicated by $(u_1, v_1)$ and $(u_2, v_2)$ $(u_1<u_2)$ in the figure, 
are locally stable points, to which $(u, v)$ tends to relax. 
Therefore, it is expected that the system behaves similar to a bistable system, where the bistability is induced by the diffusion rate asymmetry between components $U$ and $V$. 

We then consider a one-dimensional 
solution $u(x,t)$ where two stable states $u_1$ and $u_2$ are connected via a single interface. The solution expected for the steady state is of the form $u(x-ct)$; it either translates with a constant speed $c$ while maintaining the concentration profile, or stays at the same positions (i.e., $c=0$)~\cite{mori2013dissipative}. 
According to earlier studies~\cite{otsuji_mass_2007, ishihara_transient_2007, goryachev_dynamics_2008, mori_wave-pinning_2008, rubinstein2012weakly, holmes_analysis_2016, jacobs_small_2019, chiou_principles_2018, brauns2018phase, brauns2020wavelength}, the latter static solution is typical in MCRDs (see also Appendix~C). We obtain this solution when
\begin{equation}
	\Psi(u_1, p_\infty)=\Psi(u_2, p_\infty)~,
\label{eq:maxwell}
\end{equation}
is satisfied, where $\Psi(u,p)$ is defined by
\begin{equation}
\Psi(u,p) =\int^u f(u, (p-D_u u)/D_v) du~.
\label{eq:energy1}
\end{equation}
Eq.~\ref{eq:maxwell}  
determines the value of $p_\infty$ (see Appendix~C for derivation of the above relation). $u_1$ and $u_2$ corresponds to stable solutions for $d\Psi(u,p_\infty)/du=0$. 
In higher dimensions, $p_{\infty}$ obtained above is for a solution $u(\bm r)$ where two stable states are segregated by a {\it flat} interface. 
We will see that $p$ takes larger value than $p_\infty$ for curved interface, which is responsible for material transport among domains and dominates long-time coarsening dynamics. 

Examples of $\Psi(u)$ at $p=p_\infty$ are shown in Fig.~\ref{fig:fig0}d.
The double-minimum functional shape of $\Psi(u)$ seen in the figure, together with the functional form shown in Eq.~\ref{eq:energy1}, reminds us of energy functional in phase separation dynamics \cite{hohenberg_theory_1977}. Based on this analogy, we can construct 
the phase diagram of the MCRD, as shown in Fig.~\ref{fig:fig0}e. 
In the figure, the horizontal and vertical axes represent the composition of $u$ 
at the initial state and the dimensionless parameter $T=1/RD$, respectively. 
The region enclosed by the blue curve is obtained by the condition $d^2\Psi/du^2=(D_v f_u - D_u f_v)/D_v<0$, 
corresponding to the spinodal line in the phase separation.
Interestingly, this condition coincides with that for Turing instability 
shown in Eq.~\ref{eq:TuringCond}. 
Hereafter we refer to this region as the unstable region. 
The brown and black curves represent two values of $u$ that minimize $\Psi(u)$, obtained by solving $d\Psi/du=0$. These are interpreted as a binodal line. 
We call the region between the spinodal and the binodal lines as metastable region, 
where homogeneous states are stable in a deterministic sense; however, inclusion of the density fluctuation to MCRDs provokes pattern formation. 

In the following sections, we will investigate the pattern formation dynamics of MCRDs in more detail, 
separately in the unstable and metastable regions.
Through the analysis, we will test the analogy between MCRDs and phase separation phenomena.
Before proceeding to the next section, 
we make two remarks: 
first, the topology of phase diagrams depends on the specific form of the reaction term~\cite{brauns2018phase}. In Appendix D, Fig.~5b, we show a phase diagram for Model II, which does not have a parabolic shape unlike in Fig.~\ref{fig:fig0}e. 
Second, a relation equivalent to Eq.~\ref{eq:maxwell} was derived in a recent paper by Brauns et al.~\cite{brauns2018phase} in which 
the reaction-term dependence of the phase diagrams is discussed in detail. Both their and our approaches do not require any approximations, and thus the resulting phase diagram (Fig.~\ref{fig:fig0}e) improve on those presented in previous studies (see, e.g., \cite{holmes_analysis_2016,  jacobs_small_2019}).

\subsection{Pattern formation in the unstable region}
\begin{figure*}[t!]
\centering
\includegraphics[width=16.0cm]{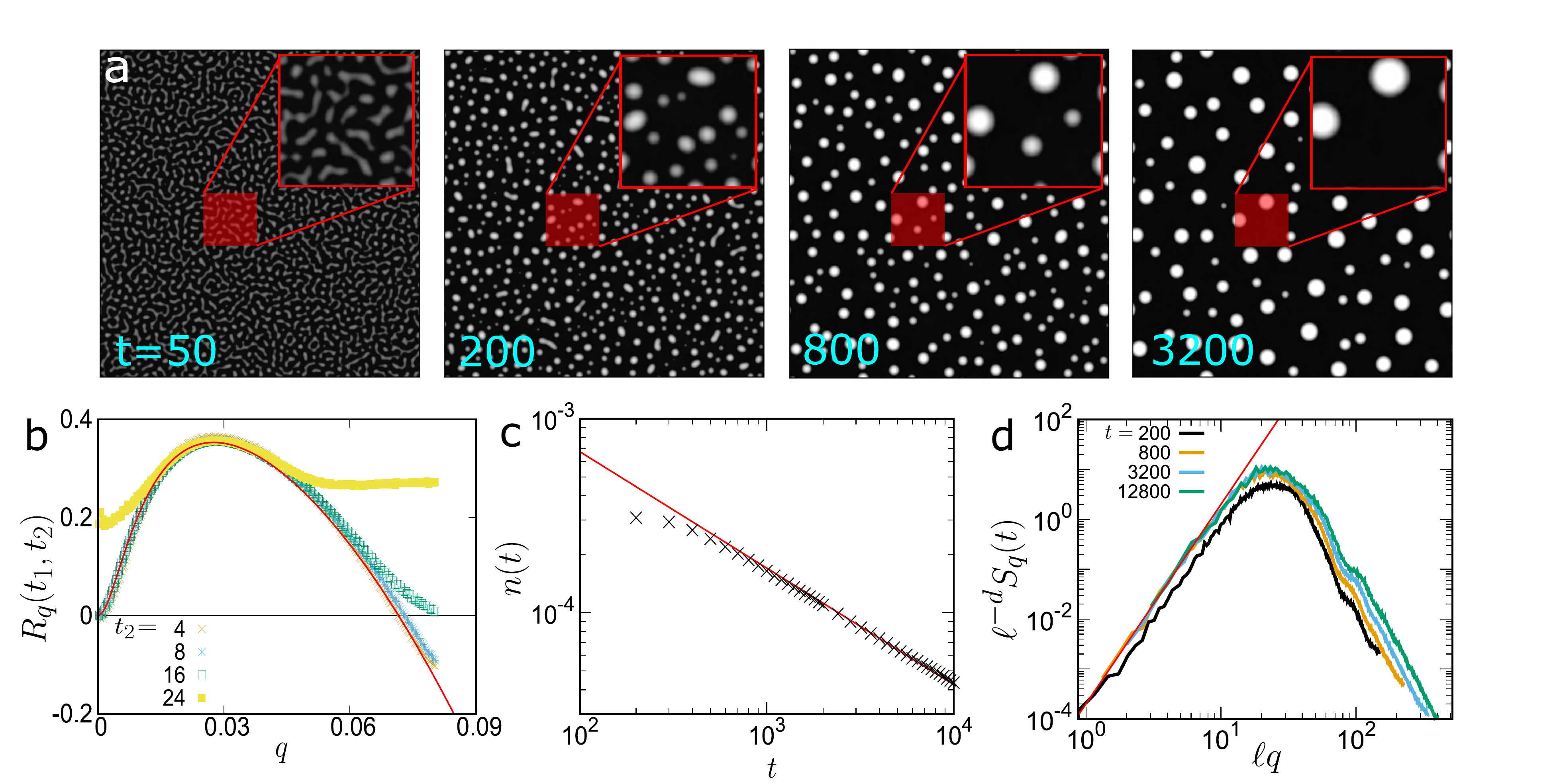}
\caption{{\bf Formation of Turing pattern in the unstable region.} 
{\bf a}. Formation process of the droplet structure. 
Spatial patterns of $s=u+v$ at $t=50,200,800$, and $3200$ are shown. The system size is $1024^2$.
The insets are enlarged images. 
{\bf b}. Growth rate measured by simulation results via a relation, $R_q(t_1,t_2)=\frac{1}{2 (t_2 - t_1)}\log(S_q(t_2)/S_q(t_1))$, where $S_q(t)$ is a structure factor at time $t$. In this figure, $t_1$ is fixed as $t_1=2$, and the results for various time $t_2$ ($= 4,8,16,24$) are shown. 
The red curves represent a theoretical prediction of growth rate in the linear regime. 
{\bf c}. Temporal changes in the number of droplets per unit volume, $n(t)$. The red line represents a power-law function with exponent $0.60$.
{\bf d}. Scaled structure factor $\ell^{-d} S(q,t)$ (where $d$ is the spatial dimension) for various time. 
Here, $\ell(t)$ is a length scale determined by $\ell^{d}n=1$, 
a typical center-of-mass distance among neighboring droplets. 
The red line represents a power function with exponent $4$. 
The above data are obtained by simulations using parameter set A indicated in Fig.~1{\bf e} 
($u_0= 3.94$, $T = 0.002)$.
}
\label{fig:fig2}
\end{figure*}
We first study the unstable region. 
To focus on similarities and differences between pattern formation in MCRDs and phase separation phenomena, in the following section, the time evolution of the conserved variable $s$ is investigated. 
Fig~\ref{fig:fig2}a shows 2D pattern formation process of $s$ for the parameter set A, indicated in Fig.~\ref{fig:fig0}e.
An initial pattern with a characteristic length spontaneously appears 
everywhere in space $(t=50)$. 
Then, elongated and branched domains are relaxed to spherical shape $(t=200)$. 
Using linear stability analysis, the deviation of $s$ from the initial homogeneous density $s_0$, $\delta s = s - s_0$, is expected to obey 
$\delta s({\bm q}, t) \sim \delta s({\bm q}, 0) \exp(-\lambda(q) t)$, where ${\bm q}$ is the wave number. $\lambda$ is the maximum growth rate of the perturbation with the wave number ${\bm q}$~(see also Fig.~\ref{fig:fig0}b), which can be estimated from the simulation results via $R_q(t_1,t_2) = \frac{1}{2 (t_2 - t_1)}\log(S_q(t_2)/S_q(t_1))$, where $S_q(t)=\langle s({\bm q}, t)s(-{\bm q}, t) \rangle$ is the structure factor at time $t$ and the bracket represents the average over the angle. If $R_q(t_1,t_2)$ is independent of time, $R_q$ is to be equivalent to $\lambda(q)$. 
Fig~\ref{fig:fig2}b shows the growth rate measured by the simulation results, 
where 
$R_q$ at different times collapse onto $\lambda (q)$ estimated by linear analysis (see red curve) for a short time. This indicates that the observed pattern formation seen is driven by the Turing instability. 

In the later stage $(t\geq400)$, 
spherical droplet-like domains coarsen over time. More specifically, the smaller droplets tend to shrink and eventually disappear, whereas the larger droplets tend to grow. 
This trend is reminiscent of the evaporation-condensation (Lifshitz--Slyozov--Wagner, or LSW) mechanism in phase separation phenomena 
as pointed out in Refs.~\cite{jacobs_small_2019, chiou_principles_2018, brauns2020wavelength}. 
To qualitatively characterize this growth behavior, we track how the number of droplets changes over time. 
Fig~\ref{fig:fig2}c shows the number density of the droplets $n(t)$, where 
$n$ shows a power-law decay for the long time regime. 
The estimated power-law exponent is approximately $0.60$ (see red line in Fig~\ref{fig:fig2}c). 
Because the volume (area) occupied by a single droplet is roughly $n^{-1}$, 
the characteristic length scale at time $t$ is estimated as $\ell = n^{-1/d}$ ($d$ being spatial dimension, i.e., $d=2$), we accordingly obtain $\ell \propto t^{-0.30}$. 
This power-law exponent, often called the growth exponent, takes a different value depending on the coarsening mechanism. The value of $0.30$ observed here is close to the growth exponent known for the LSW mechanism (1/3). 
We perform the same analysis for a 3D simulation at the same density (i.e., for parameter set A in Fig.~\ref{fig:fig0}e) and the obtained growth exponent is $0.29$. Furthermore, the obtained prefactors of the coarsening law are 9.6 and 9.8 for 2D and 3D, respectively. 
These results are consistent with the fact that coarsening law by the LSW mechanism is independent of dimensionality for $d \geq 2$. 

In phase separation phenomena, the power-law decay in the characteristic length is a consequence of a self-similar growth in the patterns. 
To examine whether this is the case for our problem, we perform a dynamic scaling analysis for $S_q$. 
In Fig~\ref{fig:fig2}d, we show that $g \equiv \ell(t)^{-d} S_q(t) $ as a function of $\ell q$. 
When a self-similar growth is the case, all the structure factors at different time 
should be mapped onto a unique mater curve. 
In this figure, we can observe that $g$ for small $q$ collapses onto a single power-law function after the formation of spherical droplets $(t \geq 400)$, indicating that the distance between center-of-mass positions of neighboring droplets are scaled by a unique length $\ell$. 
On the other side, for large $q$, 
$S_q$ slowly shifts toward the higher-$q$ side for intermediate time regime (e.g., $t\sim 3200$), meaning that the density profile of droplets is not completely scaled by $\ell$ in this time regime. 
This is because $s$ inside some droplets is not saturated to the stable fixed point $s_2$. 
We mention that such breakdown of self-similarity originating from unsaturation is not specific to MCRDs, but
is also observed in the phase separation phenomena when the shape of double-well potential is highly asymmetric (for example, see Refs.~\cite{khanna_kinetics_2010, narayanam_coarsening_2017, zhang_morphological-evolution_2017}). 

For low-$q$, $S_q$ is proportional to $q^4$, which is often observed in phase transition phenomena with a conserved order parameter (see, e.g., \cite{yeung_scaling_1988, fratzl_quenched_1989, koga_late_1993}). According to the literature, this trend appears in the case where surface tension plays a central role in the coarsening process, rather than thermal fluctuation~\cite{furukawa_multi-time_1989} and phase separating structure is statistically isotropic~\cite{tomita_preservation_1991}.

\subsection{Pattern formation in the metastable region}
\begin{figure*}[t!]
\centering
\includegraphics[width=16.0cm]{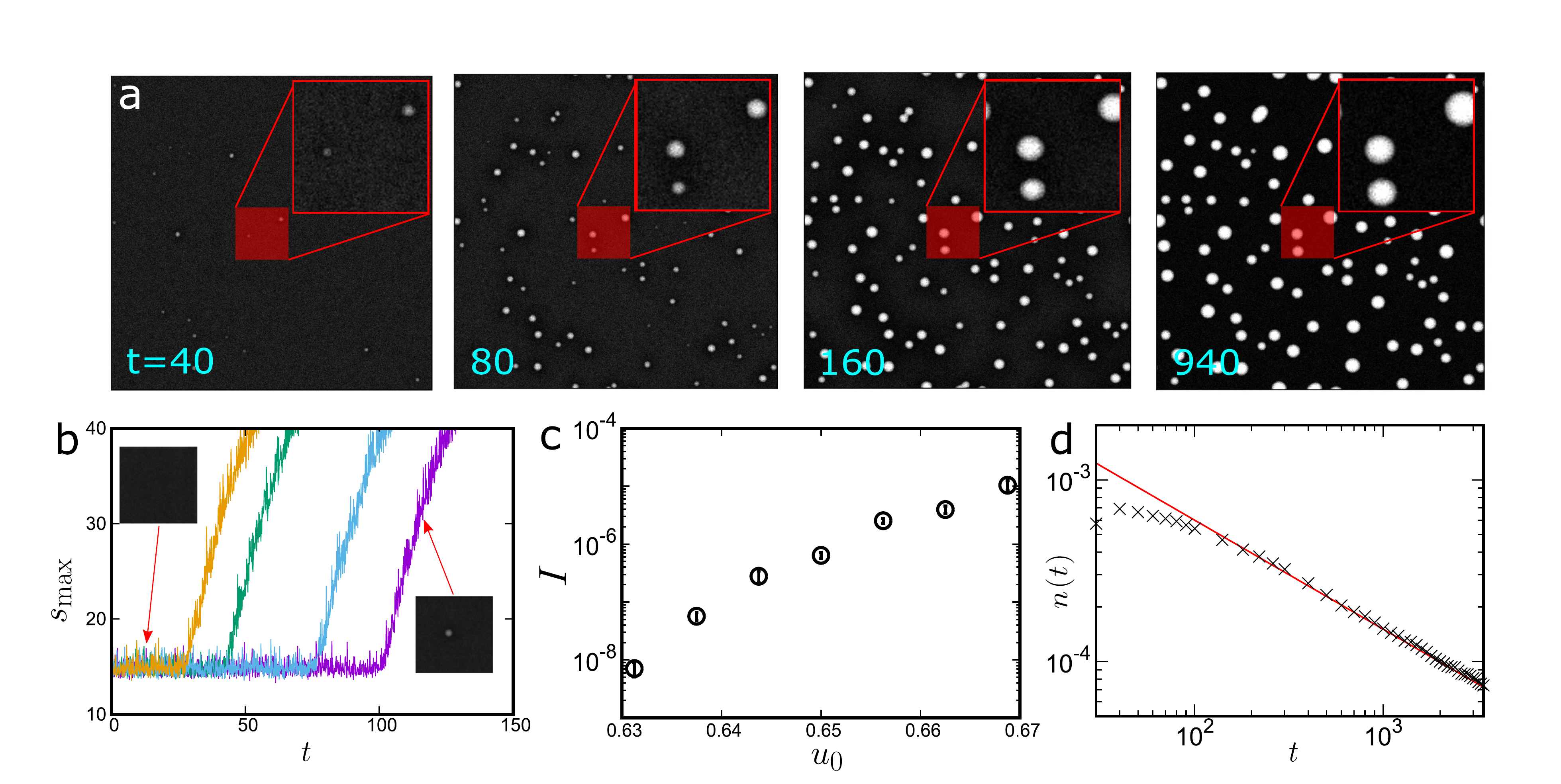}
\caption{{\bf Nucleation-growth-like pattern formation in the metastable region.} 
{\bf a}. Formation process of droplet structure for parameter set D $(u_0 = 0.65, T=0.002)$. Spatial patterns of $s = u+v$ at time $t = 40, 80, 160$, and $940$ are shown; the system size is $1295$. 
The insets are enlarged images. 
{\bf b.} Time series of the maximum value of $s$, $s_{\rm max}$, for four independent simulations using parameter set D. The sudden increase in $s_{\rm max}$ at random time indicates that the formation of the droplet is a stochastic activation process. The system size is chosen to be $162^2$, which is small enough to observe a single droplet formation in homogeneously mixed space, and large enough compared to the size of the nucleus. 
{\bf c.} Nucleation rate under various initial conditions. 
{\bf d.} Temporal changes in the number of droplets per unit volume $n(t)$ for parameter set $(u_0 = 0.70, T=0.002)$. The red line represents a power-law function with exponent $0.60$. 
}
\label{fig:fig3}
\end{figure*}
Next, we focus on pattern formation dynamics in the metastable region. 
We confirm from simulations that the homogeneous state is destabilized for parameter sets 
in the unstable region (A and B in Fig.~\ref{fig:fig0}e),
but is stable in the metastable region (C and D), which is consistent with the Turing mechanism. 
However, by choosing appropriate non-homogeneous initial conditions (while the mean density is kept constant in the metastable region), 
we find cases where systems relax to a bistable spatial structure. 
To see a typical kinetics in the metastable region, in this section 
we investigate how patterns develop in the presence of density fluctuation.
We incorporate intrinsic fluctuations originating from both diffusion and chemical reactions, referring to Ref.~\cite{biancalani_giant_2017}, see Appendix~A for details.

Fig.~\ref{fig:fig3}a shows the time evolution of conserved variable $s$. In the early stage ($t = 40, 80$), 
droplets appear at random positions and at stochastic timing. 
The size of the droplets increases with time. 
At the intermediate time ($t = 160$),
the growth of the droplets significantly slows down. 
This is because molecules in the region surrounding the droplets are almost exhausted, where $s$ is in the vicinity of the minimum stable point.
For a droplet to grow further, it is necessary to transport material from other droplets.
Indeed, at the later stage ($t=940$), the growth of the droplets is accompanied with shrinkage and eventual disappearance of small droplets. 
These pattern formation dynamics are similar to those in the nucleation-growth type phase separation of pure fluids, multi-component liquid mixtures, and alloys~\cite{kalikmanov_classical_2013}, but differs from those in so-called stochastic Turing patterns which are induced by stochastic resonance \cite{butler_robust_2009, biancalani_stochastic_2010, bonachela_patchiness_2012, butler_fluctuation-driven_2011, biancalani_giant_2017, karig_stochastic_2018}.

As an indicator of the formation of the nucleus, we monitor the time evolution of a maximum density value of $s_{\rm max}$. The results obtained by four independent simulations are shown in Fig.~\ref{fig:fig3}b. 
We can see that $s_{\rm max}$ initially fluctuates around a particular value and suddenly starts to increase. The onset time $t_{\rm inc}$ clearly indicates the birth of nucleus. 
From the onset time, we compute the nucleation rate $I$ via $I=N/\langle t_{\rm inc}\rangle V$, where $N$ is the number of nuclei at time $t_{\rm inc}$, $V$ is the volume (area), and the bracket represents the ensemble average. 
$16$ independent simulations were performed to determine $I$ under various initial conditions. 
The results are shown in Fig.~\ref{fig:fig3}c, where the vertical axis is displayed using a logarithmic scale.
The nucleation rate $I$ decreases significantly for a slight decrease in the initial density $u_0$.
Such a steep change in the nucleation rate for a small change of environment is widely observed in nucleation-type phase separation phenomena~\cite{kalikmanov_classical_2013}. 

Although the driving force of pattern formation in the early stage is entirely different between 
the unstable and the metastable regions, 
we observe that the coarsening behavior in the late stage are very similar for these regions; 
small droplets evaporate while large ones grow. 
This similarity is quantitatively confirmed by plotting $n(t)$, the temporal change in the number density of droplets
(Fig.~\ref{fig:fig3}d).
We find the same trend $\ell \propto t^{-0.30}$ as in Fig.~\ref{fig:fig2}c, 
further supporting that the LSW mechanism can be applied to MCRDs.

\subsection{Coarsening process in the late stage}
\begin{figure*}[t!]
\centering
\includegraphics[width=16.0cm]{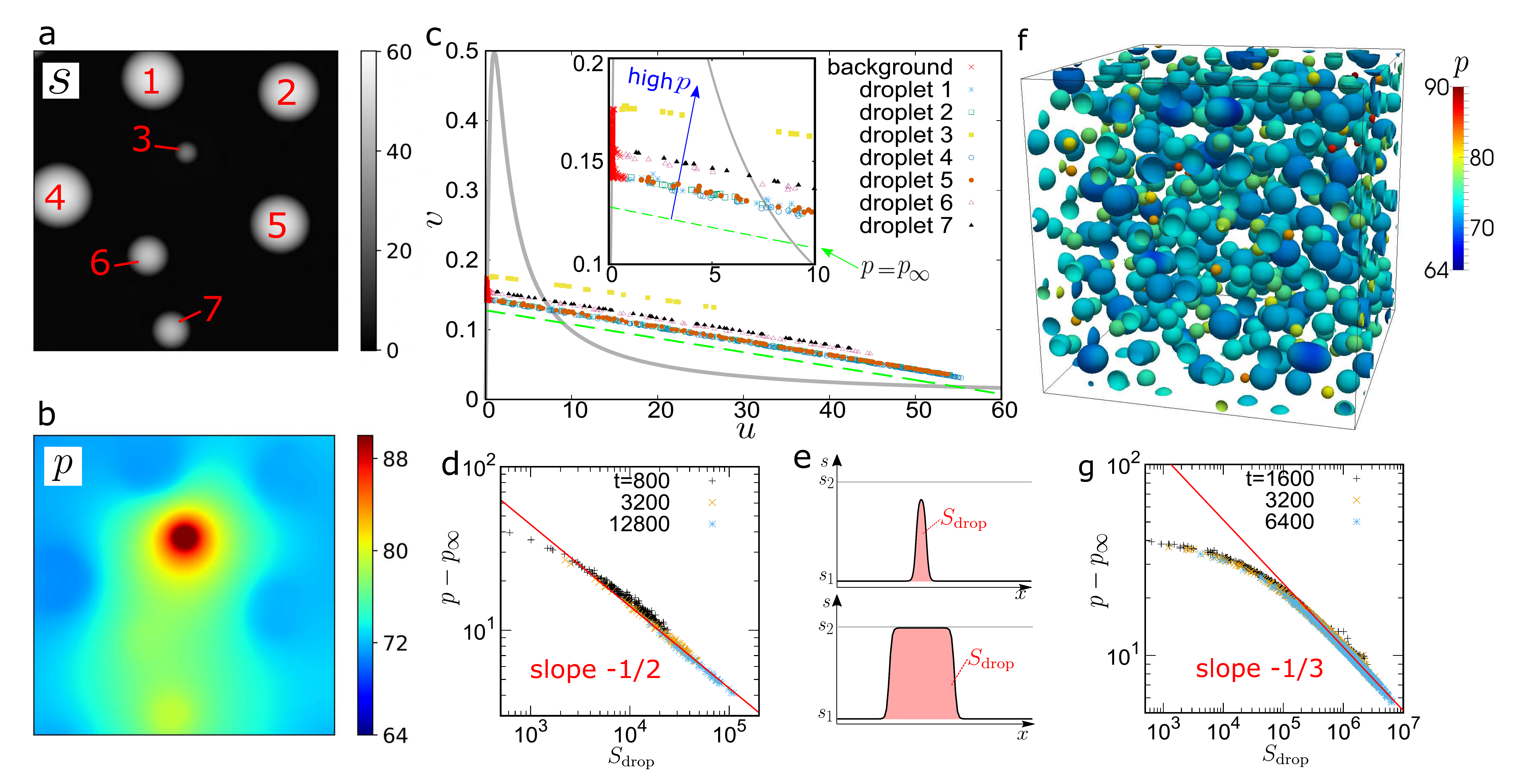}
\caption{{\bf Evaporation-condensation process in the coarsening regime.} 
{\bf a}. Density profile of conserved variable $s$ at time $t=3200$ at parameter set A in Fig.~1{\bf e}. 
{\bf b}. Spatial distribution of ``chemical potential'' $p$ corresponding to panel {\bf a}.  
The image size is $256^2$.
{\bf c}. Mapping the states of volume elements on the $u$-$v$ space for panels {\bf a} and {\bf b}. 
Different symbols indicate to which droplets the volume elements belong (see the numbers labeled on the droplets in panel {\bf a}). 
When a volume element does not belong to any droplets, it is referred to as ``background''.
The gray curve is the nullcline of the reaction term ($f(u,v)=0$). The green dashed line is $p_\infty = D_u u + D_v v$, where $p_\infty$ is the value of $p$ in the limit where the droplet size is sufficiently large (see below). 
{\bf d}. Droplet size dependence of $p$. Mean values of $p$ for each droplet, subtracted by $p_\infty$, are plotted against the droplet size $S_{\rm drop}$. Data are sampled at three different times. 
$p$ is expected to be equal to $p_\infty$ for sufficiently large droplet
($p_{\infty} \simeq 64$ in the current parameter setting).
Droplet size $S_{\rm drop} =\int (s-s_1) d{\bm r}$ indicates an excess amount of $s$ from the minimum stable point $s_1$,
where the integral is taken over each droplet. The red line is a power function with exponent $-1/d$ (where $d$ is the spatial dimension). 
{\bf e}. The relation between droplet size $S_{\rm drop}$ and density profile $s$. 
For small $S_{\rm drop}$, the maximum value of $s$ cannot reach to the largest stable point $s_2$ (top).
For a sufficiently large $S_{\rm drop}$, the most of the body elements in the droplet have a value close to $s_2$, and the interface connecting the inside and outside of the droplet has an almost constant width (bottom). 
{\bf f}. A spatial pattern obtained from a 3D simulation for time $t=6400$ at parameter set D. 
A contour surface with $s=20$ is shown. The color is labeled according to the value of $p$ at the contour surface. 
The system size is $1024^3$.
{\bf g}. $S_{\rm drop}$-$p$ mapping obtained from 3D simulation.  
}
\label{fig:fig4}
\end{figure*}
Here, we investigate the evaporation-condensation (LSW) process observed in both the unstable and metastable regions in more detail.
Since the time evolution of the conserved variable $s$ proceeds following the gradient of $p=D_u u + D_v v$ (see Eq.~\ref{eq:conseved}), the variable $p$ is responsible for the transport of $s$ among the droplets. 
Figs.~\ref{fig:fig4}a and b show an enlarged image of Fig.~\ref{fig:fig2}a at $t=3200$ and the corresponding spatial map of $p$, respectively. We can see that smaller droplets have a higher $p$ value. In this sense, we may call $p$ as ``chemical potential'' of $s$, on the analogy of phase separation phenomena~\cite{onuki2002phase}. 

To verify the validity of this physical interpretation in a quantitative manner, we divide the image into $64^2$ equal-size square sections and map the states of the sections onto the $u$-$v$ space. The results are shown in Fig.~\ref{fig:fig4}c. 
Each point in this figure corresponds to the state of each section. 
The symbols of the points represent the droplets to which the sections belong (see the number labeled on the droplets in Fig.~\ref{fig:fig4}a). When a section is located outside the droplets, we use a red cross symbol
(`background' in the figure). 
We determine whether a section is inside or outside of the droplets by whether $s$ at that section is above or below a threshold value $s_{\rm th} = 2.2$, which is slightly higher than the smaller stable point $s_1$. In this figure, we can see that points in the same droplet are distributed on a line with constant $p$, and smaller droplets have higher $p$. 
This result indicates that ``chemical potential'' $p$ is uniquely determined by the morphology of the droplet, which is suggestive of the existence of the surface tension. 
We also confirm similar results for the 3D simulation and the 2D simulation of Model II (see Fig.~\ref{fig:fig4}f and Appendix~D, Figs.~5 and 6, respectively). 

To analyze these results in further detail, in Fig~\ref{fig:fig4}d 
we show the simultaneous distribution 
of droplet size, $S_{\rm drop}=\int (s-s_1) d{\bm r}$ (the integral is taken within each droplet) and the value of $p$ in the droplets. The data are sampled at three different points in time ($t = 800$, $3200$, and $12800$). 
The results clearly show that there is a one-to-one correspondence between $S_{\rm drop}$ and $p$. Furthermore, for the large $S_{\rm drop}$ region we find the relation $(p-p_{\infty}) \propto S_{\rm drop}^{-1/d}$ where $d$ is the spatial dimension (see, Fig.~\ref{fig:fig4}d). 
This asymptotic behavior can be understood as follows: when $S_{\rm drop}$ is large, 
$s$ in the droplet saturates to the larger stable point $s_2$ and is connected to 
the outside of the droplet via the interface, the width of which is almost constant and independent of the droplet size.
(see the bottom panel of Fig~\ref{fig:fig4}e). In this situation, 
$S_{\rm drop}$ can be approximated as $S_{\rm drop} \sim (s_2-s_1) V_{\rm drop}$, 
where $V_{\rm drop}$ is the volume (area) of the droplet.
Since the droplets have a spherical shape, by denoting the radius as $R$, 
we can obtain the relationship $(p-p_{\infty}) \propto 1/R$. This is the same form as the Young--Laplace equation, where the proportionality constant of 
the above relation is expected to be associated with the surface tension in MCRDs.
Note that $p$ coincides with $p_\infty$ for a large curvature limit ($R\rightarrow \infty$), a situation in which two stable states are segregated by a flat interface, as discussed In Sec.~III~A.
The same relationship is confirmed for the 3D simulation (see Fig~\ref{fig:fig4}g) and a 2D simulation with a different reaction term (Model II, see Appendix D, Fig~7c).
When $S_{\rm drop}$ is small, $s$ in the droplets is far from $s_2$ and the above approximation does not hold (see the top panel of Fig~\ref{fig:fig4}e), leading to a deviation of $p-p_{\infty}$ from $\propto S_{\rm drop}^{-1/d}$. 
The reason why the density profile of the droplets changes depending on the droplet size is that initial density $s_0$ is very far from the stable fixed point $s_2$; thus, it takes long time for $s$ to saturate to $s_2$ while retaining the mass conservation law for $s$. 
We note again that such a trend is not a feature specific to MCRDs, but is also seen in phase separation phenomena with a highly asymmetric shape in the free energy form \cite{khanna_kinetics_2010, narayanam_coarsening_2017, zhang_morphological-evolution_2017}.

\subsection{Reduced mass transport equation}
Finally, we discuss the physical origins of ``surface tension'' in MCRDs. 
This can be translated into a question of whether or not the variable $p$ that we call 
``chemical potential'' can be written in some variational form. 
The governing equation of $p$ is
\begin{equation}
\frac{\partial p}{\partial t} = (D_u + D_v) \nabla^2 p - D_u D_v \nabla^2 s + (-D_u+D_v) f. 
\label{eq:pTime}
\end{equation}
Eq.~\ref{eq:conseved} and \ref{eq:pTime} provide a closed set of partial differential equations 
with respect to $s$ and $p$, which is equivalent to the original equation for MCRDs (Eq. \ref{eq:RD1_1} 
and \ref{eq:RD1_2}).

Since variable $s$ is a conserved quantity, the time evolution of $s$ is 
driven only by the diffusive transportation of molecules (see Eq.~\ref{eq:conseved}) and
is expected to proceed slowly compared with that of $p$ (Eq.~\ref{eq:pTime}). 
Indeed, in the simulations we observed that the time derivative of $p$ quickly decays in the early linear time regime and $\partial p/\partial t$ is close to $0$, 
where the terms on the right-hand side of Eq.~\ref{eq:pTime} are balanced. Thus, $p$ is subjective to $s$, i.e., we can regard $p$ as a functional of $s$. 
The validity of this description can be confirmed by a one-to-one correspondence between $p$ and $s$, as shown in Figs.~\ref{fig:fig4}d and g. 
In addition, since $p$ at steady states becomes spatially constant,
the spatial profile of $p$ is gradual in the coarsening regime (see Fig.~\ref{fig:fig4}b).
Thus, $\nabla^2 p$ can be regarded as a small perturbative term and we can expand $p$ as
\begin{equation}
	p[s] = P(s) +  \mathcal{A}(s) \nabla^2P(s)~
\label{eq:reductionEq}
\end{equation}
where we neglect the spatial derivatives higher than the forth order (see Appendix~E for detailed derivation and precise meaning of the above expression). 
$P(s)$ provides the leading approximation, which is given by a steady-state solution of $p$.
$\mathcal{A}(s)$ is also a function of $s$ that is determined by bistable steady stable solution. 
Substituting Eq.~\ref{eq:reductionEq} into Eq.~\ref{eq:conseved}, we obtain a mass transport equation in the following form:
\begin{equation}
   \label{eq:reductionEqFull}
	\frac{\partial s}{\partial t} = \nabla^2 \left[ P(s)-\kappa(s) \nabla^2 s + \zeta(s) (\nabla s)^2 \right]~,
\end{equation}
where $\kappa(s)$ and $\zeta(s)$ are defined as
$\kappa(s) \equiv -\mathcal{A}(s)dP/ds$ and $\zeta(s) \equiv \mathcal{A}(s)d^2P(s)/ds^2$. 
This equation for mass transport satisfies generic conditions such as mass conservation and spatial inversion symmetry of the system. 
Notably, Eq.~\ref{eq:reductionEqFull} has the same form as the coarse-grained equation obtained for the system showing motility induced phase separation (MIPS), for which Solon et al. showed that the system can be mapped into a variational form~\cite{solon_generalized_2018}.
This indicates that Eq.~\ref{eq:reductionEqFull} also follows the same variational principle.

At the lowest order, $s$ obeys the non-linear diffusion equation $\partial s/\partial t \simeq \nabla \left[(dP/ds) \nabla s \right]$, indicating that the sign of ``diffusion coefficient'' $dP(s)/ds$ determines the stability of the homogeneous solution. 
In general, one can prove the relationship $dP(s)/ds=(D_v f_u-D_u f_v)/(f_u - f_v)$, and $dP/ds <0$ is nothing but Turing instability condition (see also the growth rate $\lambda(q)$ in the caption of Fig.~\ref{fig:fig0}b).  
For Model I, we can analytically evaluate $P(s)$ and obtain the same stability condition  
$T<T_c=1/8$ as before (Appendix~F). 
These consistencies support the validity of the above reduction. 
\\

\section{Conclusion}
In summary, we find that the pattern dynamics in MCRDs, starting from a uniform state towards an eventual single isolated domain, are classified into two types, similar to phase separation phenomena~(see also Ref.~\cite{brauns2018phase}). 
One is triggered by the absolute instability of the initial uniform state and the other goes through a nucleation and growth process. 
This classification is systematically addressed via a phase diagram, which can be constructed by the derivative of a free-energy like function introduced by Eq.~\ref{eq:energy1}.
Furthermore, we performed large-scale numerical simulations in both two and three dimensions. 
We confirmed that in the late stage, droplet patterns appearing both in stable and metastable regions obey a relation similar to the Yong-Laplace equation, and their coarsening process is limited by the evaporation-condensation mechanism with growth exponent $1/3$. 
These results suggest that in the presence of conserved variables, a surface-tension-like quantity generally emerges 
in RDs. 
By a reduced description of the systems focusing on the dynamically slow mode associated with mass conservation, we show that the origins of ``surface tension'' can be understood from chemical potential like functional $p[s]$ following a variational form.

MCRDs were originally introduced and have been discussed as models for molecular localization such as membrane-bounded GTPases which are responsible for the formation of cell polarity~\cite{otsuji_mass_2007, ishihara_transient_2007, mori_wave-pinning_2008, goryachev_dynamics_2008, rubinstein2012weakly, holmes_analysis_2016, chiou_principles_2018, jacobs_small_2019, brauns2018phase, brauns2020wavelength} and neuron dendrites~\cite{chen_phase_2020}, and phosphoinositide lipids~\cite{hansen_stochastic_2019}.
Recently, MCRDs have been also applied to various systems such as oscillatory motion in min-protein~\cite{frey_protein_2018} 
and chemical turbulence~\cite{halatek_rethinking_2018}, by which the importance of MRCDs for pattern formation in biological systems is further recognized. 
Although the details of chemical reactions are significantly different among the systems,
MCRDs show universal dynamics and can be relevant for many existing systems. 
One interesting possible application of MCRDs is for liquid-liquid phase separation observed inside cells \cite{hyman2014liquid, brangwynne2015polymer, boeynaems2018protein, berry2018physical}.
Liquid droplets found in cells contain many chemical components that regulate each other; 
thus, some droplets may be formed by chemical reactions consuming cellular energy rather than non-molecular specific physical interactions among proteins. 
MCRDs can provide a natural framework to model and understand the general mechanism of such phenomena.

Although MCRDs exhibiting pattern formations do not satisfy detailed balance, they show very similar behavior with phase separation driven by the thermodynamic variational principle. In this sense, phase separation observed in MCRDs can be regarded as a non-equilibrium extension of the typical phase separation phenomena. It is then interesting to compare MCRDs with MIPS, another non-equilibrium extension of phase separation, where self-propelled particles with repulsive interaction
spontaneously form spatially localized assemblies~\cite{cates2015motility, gompper20202020}.
In the systems showing MIPS, transport of molecules is supposed to be active, 
while in MCRDs transport itself is by simple diffusion and activity is introduced into chemical reactions. 
Despite this difference, both out-of-equilibrium systems show similar dynamics of phase separation,
indicating that one could find a theoretical principle to understand both phenomena
in an integrated manner.
For MIPS, Solon et al. derived a time evolution equation for particle density, and based on
the equation they proposed a generalized variational principle, as a generalization of
the variational principle of equilibrium system \cite{solon_generalized_2018}. 
For MCDRs, the reduction equation Eq.~\ref{eq:reductionEqFull} that we derived in the present study is of the same form as that derived for the particle density in MIPS. Therefore, the generalized variational principle could be applicable to MCRDs.
Although an energetic interpretation of the extended variational functionals is still lacking at present, 
the study of this common mathematical structure is one of directions to be explored in the future. 

\section*{ACKNOWLEDGMENTS}
Numerical simulations were performed on the SGI ICE XA/UV hybrid system at the ISSP at the University of Tokyo. 
This study was supported by 
JSPS KAKENHI (JP20K14424 and JP18057992) and JST CREST JPMJCR1923, Japan.

\appendix

\section{MCRD with fluctuations}
The mass-conserved reaction-diffusion equation including fluctuations is given as follows: 
\begin{eqnarray}
\partial_t u & =  D_u \nabla^2 u - f(u,v) + \nabla \cdot {\bm j}_u - \eta_r,  \label{eq:Uequation} \\
\partial_t v & =  D_v \nabla^2 v + f(u,v) + \nabla \cdot {\bm j}_v + \eta_r. \label{eq:Vequation}
\end{eqnarray} 
Here $D_u$ and $D_v$ are the diffusion coefficients of chemical species, $U$ and $V$, and we assume $D_u < D_v$. $f$ is the reaction term. 
${\bm j}_u$ and ${\bm j}_v$ are fluctuations accompanying with diffusion of $U$ and $V$, respectively, satisfying the below statistics,
\begin{align*}
&\langle j_{u, \alpha}(t,{\bm r}) j_{u, \beta}(t',{\bm r'})\rangle = 2D_{u}u\delta(t-t')\delta({\bm r}-{\bm r'}) \delta_{\alpha,\beta}, \\
&\langle j_{v, \alpha}(t,{\bm r}) j_{v, \beta}(t',{\bm r'})\rangle = 2D_{v}v\delta(t-t')\delta({\bm r}-{\bm r'}) \delta_{\alpha,\beta}. 
\end{align*}
$\eta_r$ is fluctuation accompanying with reaction and satisfies
\begin{align*} 
\langle \eta_r(t,{\bm r}) \eta_r(t',{\bm r'})\rangle  = \sum_i|{\rm Reaction}_i|\delta(t-t')\delta({\bm r}-{\bm r'}). \label{eq:Ffluct}
\end{align*}
In our study, we use a reaction term of the form $f = f_+ - f_-$ where $f_+ = k_u u$ and $f_- = k_v v$. In this case, $\sum_i|{\rm Reaction}_i| = f_+ + f_-$.

\section{Non-dimensionalization}
By denoting the time and space units as $\tau$ and $\ell$, respectively, and taking the units for $u$, $v$ and $f$ as $u_0$, $v_0$ and $f_0$, respectively, we can rewrite Eq.~\ref{eq:Uequation} and \ref{eq:Vequation} as,
\begin{eqnarray}
\frac{\partial}{\partial \tilde{t}} \tilde u =& \tilde{\nabla}^2 \tilde{u} - R \tilde{f} + \sqrt{R} \sigma \tilde{\nabla} \cdot \tilde{\bm j}_u - R\sigma \tilde{\eta}_r, \\
\frac{\partial}{\partial \tilde{t}} \tilde v =& D \tilde{\nabla}^2 \tilde{v} + \tilde{f} + \sqrt{D} \sigma \tilde{\nabla} \cdot \tilde{\bm j}_v + \sigma \tilde{\eta}_r, 
\label{eq:RD1}
\end{eqnarray}
where we define $R=v_0/u_0$, $D=D_v/D_u$, $\sigma=1/\sqrt{v_0 \ell^d}$, $\ell = \sqrt{D_u \tau}$, and $\tau=v_0/f_0$. $\tilde{A}$ represents scaled variable of $A$. 
The statistics of the noise terms are then given as $\langle \tilde{j}_{u, \alpha}(\tilde{t},\tilde{{\bm r}}) \tilde{j}_{u, \beta}(\tilde{t}',\tilde{{\bm r}}')\rangle = 2\tilde{u} \delta(\tilde{t}-\tilde{t}')\delta({\bm r}-{\bm r'}) \delta_{\alpha,\beta}$, 
$\langle \tilde{j}_{v, \alpha}(\tilde{t},\tilde{{\bm r}}) \tilde{j}_{v, \beta}(\tilde{t}',\tilde{{\bm r}}')\rangle = 2\tilde{v} \delta(\tilde{t}-\tilde{t}')\delta(\tilde{\bm r}-\tilde{\bm r}') \delta_{\alpha,\beta}$ 
and $\langle \tilde{\eta}_r(\tilde{t},\tilde{{\bm r}}) \tilde{\eta}_r(\tilde{t}',\tilde{{\bm r}}')\rangle = (\tilde{f}_+ + \tilde{f}_-) \delta(\tilde{t}-\tilde{t}')\delta(\tilde{\bm r}-\tilde{\bm r}') \delta_{\alpha,\beta}$.
From the above equations, we can immediately obtain the time evolution equation for the conserved variable $\tilde{s} = \tilde{u} + R \tilde{v}$ as, 
\begin{eqnarray}
\tilde \partial_t \tilde s =& \tilde{\nabla}^2 \tilde{p} + \sqrt{R} \sigma \tilde{\nabla} \cdot \tilde{\bm j}_p,
\label{eq:con}
\end{eqnarray}
where $\tilde{p}=\tilde{u}+RD\tilde{v}$. Here we use $u_0$ as a common unit for $s$ and $p$. The last term on the right-hand side 
represents the diffusion noise that satisfies $\langle \tilde{j}_{p, \alpha}(\tilde{t},\tilde{\bm r}) \tilde{j}_{p, \beta}(\tilde{t}',\tilde{\bm r}')\rangle = 2\tilde{p} \delta(\tilde{t}-\tilde{t}')\delta(\tilde{{\bm r}}-\tilde{\bm r}') \delta_{\alpha,\beta}$. 

For Model I ($k_u=ca/(b+u^2)$, $k_v=c$), we choose $u_0=\sqrt{b}$, $v_0=a/\sqrt{b}$ and $f_0=ac/\sqrt{b}$, and the resulting reaction term is $\tilde{f} = \tilde{u}/(1+\tilde{u}^2)-\tilde{v}$. 
For Model II ($k_u=\alpha$, $k_v=k_0 + \beta u^2/(K^2+u^2)$), we set $u_0=K$, $v_0=\alpha K/\beta$, $f_0=\alpha K$ and $k_0=0$, and the resulting reaction term is
$\tilde{f} = \tilde{u}^2 \tilde{v}/(1+\tilde{u}^2)-\tilde{u}$. 

In the absence of noise terms ($\sigma=0$), bistable points in the steady-state solution, ($\tilde{u}_1$, $\tilde{v}_1$) and ($\tilde{u}_2$, $\tilde{v}_2$), are obtained using two equalities $\tilde{f}=0$ and $\tilde{u}+RD\tilde{v}=\tilde{p}_\infty$ (see Fig.~1{\bf c}), where $\tilde{p}_\infty$ is determined by Eq.~5 (see also Appendix~C). 
Because $\tilde{f}$ is parameter-free in our normalization, the existence of such a solution depends solely on $RD \equiv 1/T$ as shown in Fig.~1{\bf e}. Note that 
if we allow $k_0$ to be varied freely for Model II,
another parameter needs to be considered in addition to $T$. 

\begin{figure}[b]
\centering
\includegraphics[width=8.5cm]{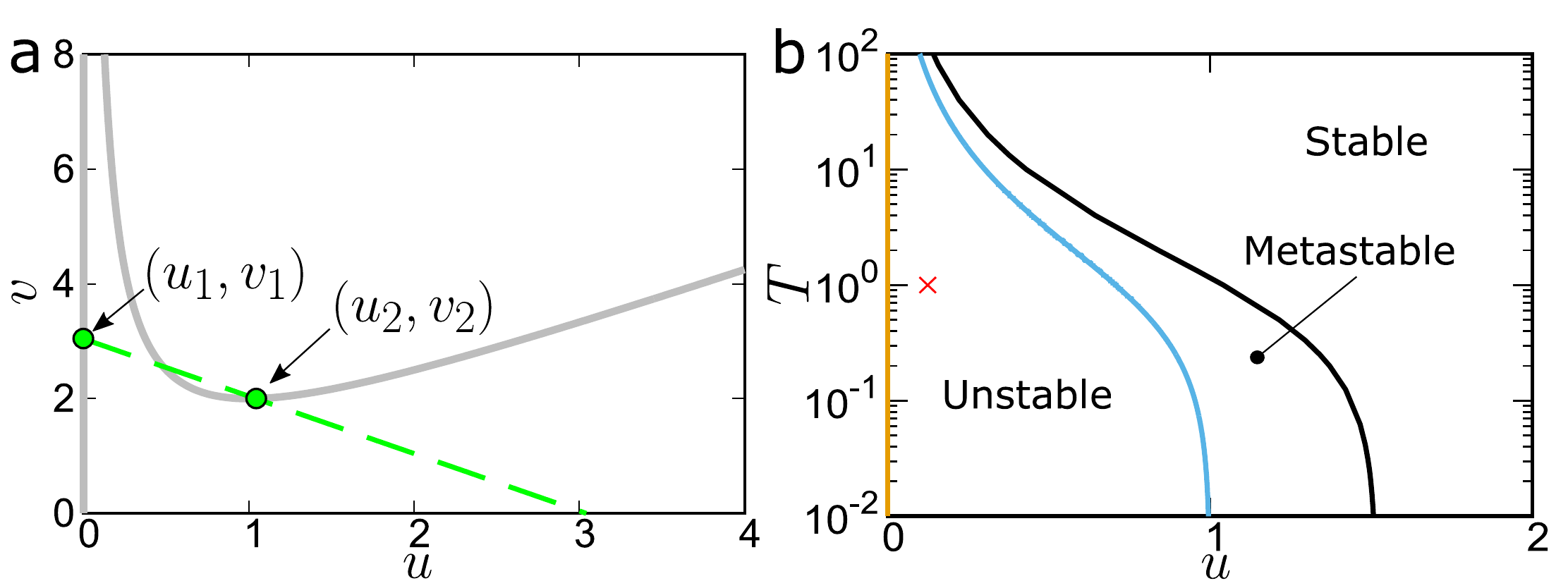}
\caption{{\bf A steady-state analysis for Model II.} 
{\bf a}. Mapping of the steady stable points on the $u$-$v$ space for $T=1$. The gray curve is the nullcline for the reaction term $f$. The green dashed line represents $p_{\infty} = D_u u + D_v v$. The green circles indicate the stable fixed points. 
{\bf b}. Phase diagram. The blue curve represents the border of the unstable region where the Turing mechanism is operative. 
The black and gray curves are the set composed of the steady stable points, $u_1$ and $u_2$, respectively (see the green circles in panel a). Here we focus on the pattern formation process at the red cross point $(u=0.125, T=1)$. 
}
\label{fig:figS1}
\end{figure}

By the non-dimensionalization described above, the condition for the Turing instability (Eq.~4) is rewritten as $ \tilde{f}_{\tilde{u}} - T \tilde{f}_{\tilde{v}} < 0$. 
For Model I, $T<1/8 ~(\equiv T_{\rm c}$) is necessary for this inequality to be satisfied and $T_{\rm c}$ is the value of $T$ at the vertex of a parabolic function shown in Fig.~1{\bf e}. Such an upper bound does not exist for Model II (see Fig.~\ref{fig:figS1}{\bf b}).

All the data presented in the figures are normalized by the units introduced here.

\begin{figure*}[t]
\centering
\includegraphics[width=11.5cm]{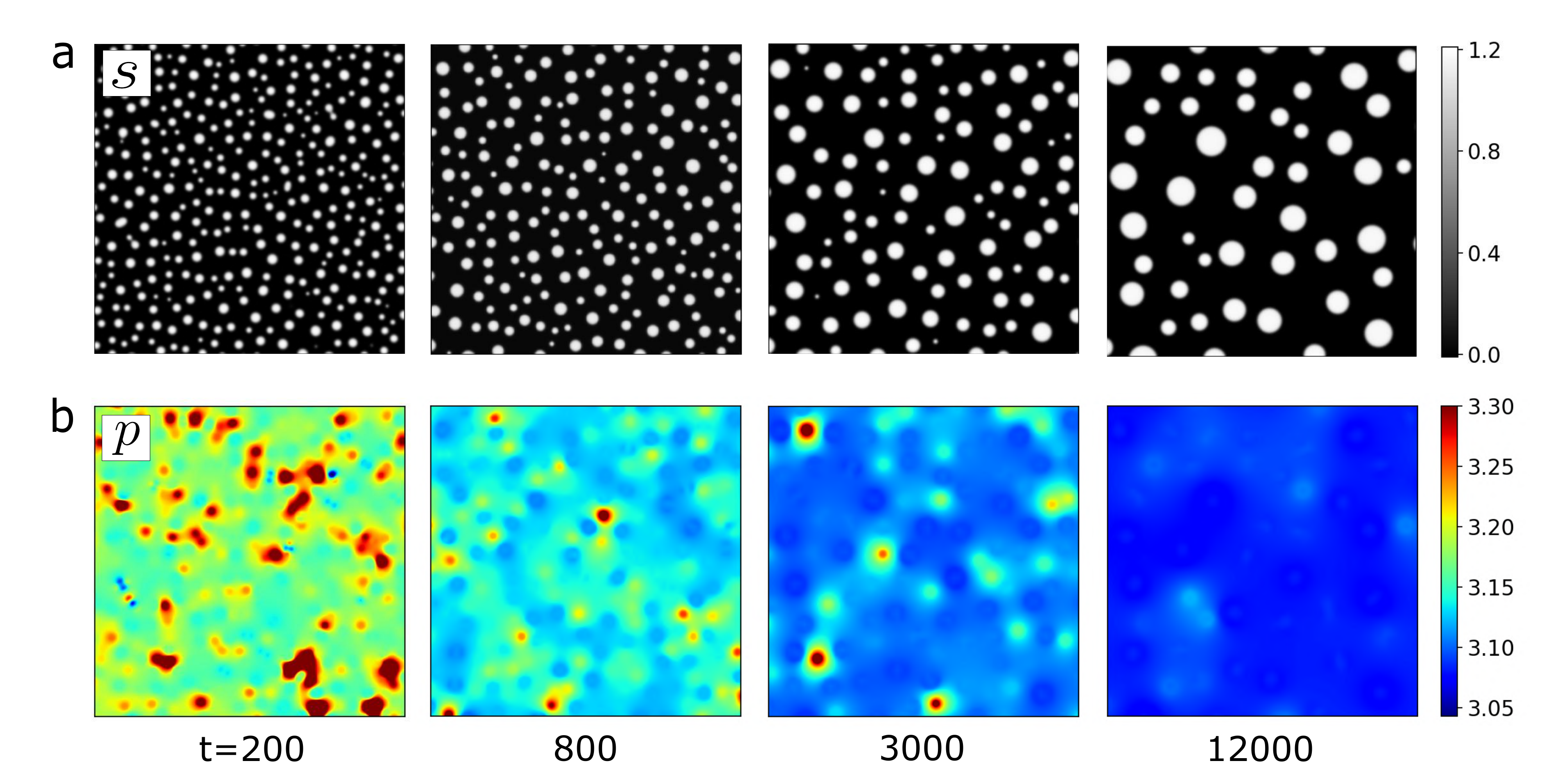}
\caption{{\bf Temporal change of the pattern at $(u, T)=(0.125, 1)$.} 
Panels {\bf a} and {\bf b} show the temporal changes in the field variables $s$ and $p$, respectively. The side length of the images is $1024$. The minimum value of the color bar for $p$ is set to $p=p_{\infty} \sim 3.04$.
}
\label{fig:figS2}
\end{figure*}

\begin{figure*}[t]
\centering
\includegraphics[width=14.5cm]{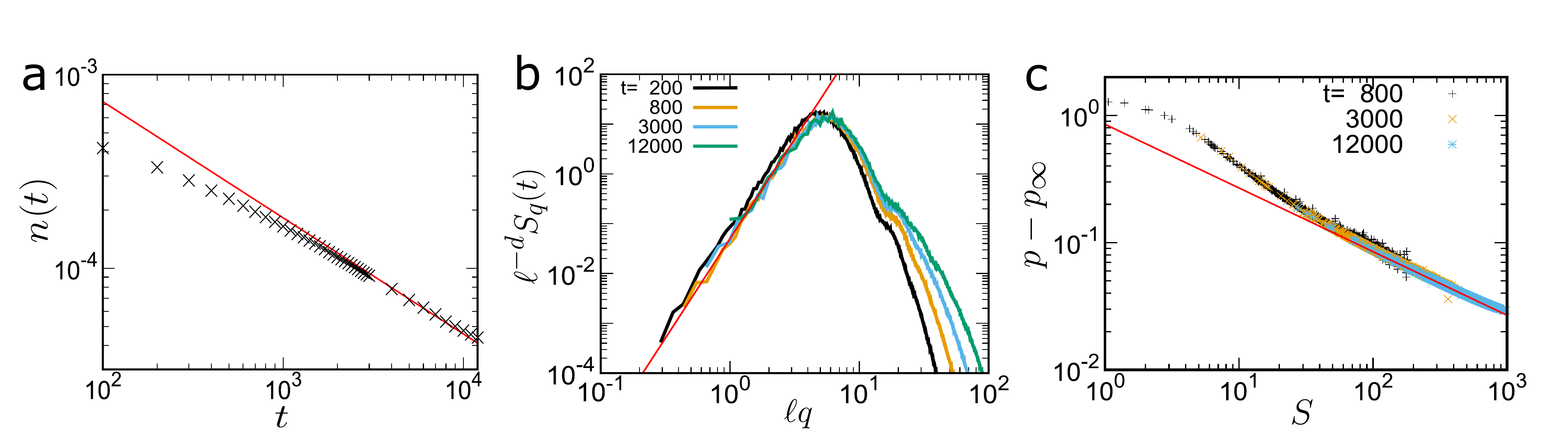}
\caption{{\bf An analysis of the coarsening process at $(u, T)=(0.125, 1)$.} 
{\bf a}. Temporal changes of the number of droplets per unit volume $n(t)$. The red line represents a power-law function with a slope of 0.60. 
{\bf b}. Scaled structure factor $\ell^{-d} S(q,t)$ (where $d$ is the spatial dimension) at various time. $\ell(t)$ is a length scale determined by $\ell^{d}n=1$, which approximately corresponds to a typical distance between neighboring droplets. The red line represents a power function with a slope of 4. 
{\bf c}. Droplet size ($S$) dependence of ``chemical potential'' $p$. 
The red line is a power function with an exponent of $-1/d$ ($d$ being the spatial dimension; $d=2$ in the current case). 
}
\label{fig:figS3}
\end{figure*}

\section{Derivation of Eq.~5}
We consider a solution for MCRDs in one dimension, where $(u,v)$ approaches 
$(u_1,v_1)$ and $(u_2,v_2)$ for $x \rightarrow - \infty$ and $+\infty$, respectively. 
In a steady condition, Eq.~3 becomes $\nabla^2 p = 0$ and the general solution is given by 
$p = Ax+B$, where $A$ and $B$ are constants. 
If $A$ is nonzero, $p$ (and thus, at least one of $u$ or $v$) diverges for $x \rightarrow \pm \infty$, which is unphysical. Thus, $p$ has a constant value, which we denote as $p_{\infty}$. 
As discussed in the main text, we consider a solution that propagates with a constant velocity $c$~\cite{mori2013dissipative}. Substituting $u(x,t)=u(x-ct)$ into Eq.~1, we obtain the following relationship
\begin{eqnarray}
-c\frac{du}{dz} = D_u \frac{d^2 u}{dz^2} - f(u, \frac{p_\infty - D_u u}{D_v} ), 
\end{eqnarray}
where $z=x-ct$ is a moving coordinate at speed $c$. By multiplying the above equation by $du/dz$ and taking integral over the entire space, we obtain
\begin{equation}
\begin{split}
-c \int_{-\infty}^{\infty} \left(\frac{du}{dz}\right)^2 dz
= \int_{u_1}^{u_2} \frac{D_u }{2}\frac{d}{dz}\left( \frac{du}{dz}\right)^2 du \\
- \int_{u_1}^{u_2} f(u, \frac{p_\infty - D_u u}{D_v} ) \frac{du}{dz}du.
\end{split}
\end{equation}
The first term on the right-hand side vanishes because $d u/dz = 0$ for $z=\pm \infty$.  
Therefore, $c$ is related to $u(z)$ as follows: 
\begin{eqnarray}
c= \left[ \Psi(u_2, p_\infty) - \Psi(u_1, p_\infty) \right] / \int_{-\infty}^{\infty} \left(\frac{du}{dz}\right)^2 dz. 
\end{eqnarray}
Here, $\Psi$ is an integral of $f$ with respect to $u$ introduced in Eq~6.
At the stationary state where $c = 0$,
two stable states are related by $\Psi(u_2, p_\infty) = \Psi(u_1, p_\infty)$.
This equality determines a position-independent value of $p = p_\infty$.
Note also that $u_1$ and $u_2$ are two minima of $\Psi(u, p_\infty)$, satisfying 
 $d\Psi(u ,p_\infty)/du=f(u,(p_\infty-D_u u)/D_v)=0$.
In the main text, $p_{\infty}$ is evaluated by assuming the periodic boundary conditions for which the system size $L$ is sufficiently large.

The existence of the static stationary solution can be confirmed by introducing a functional defined as $\mathscr{E} [u,p] \equiv \int d{\bm r} [ \Psi(u,p) + \frac{D_u}{2} |\nabla u|^2 ]$.
The functional derivative of $\mathscr{E}$ with respect to $u$ relates to the time derivative of $u$ as $\partial u/\partial t = - \delta \mathscr{E}/ \delta u$. 
Under the condition of constant $p=p_{\infty}$, $\mathscr{E}[u,p_\infty]$ acts as a Lyapunov function because
\begin{eqnarray}
\frac{d\mathscr{E}[u,p_{\infty}]}{dt} = - \int \left(\frac{\partial u}{\partial t}\right)^2 d{\bm r}
 \leq 0 ~.
\end{eqnarray}
If we assume a traveling solution of the form $u(x-ct)$,
$d\mathscr{E}[u,p_{\infty}]/dt = -c^2 \int (du/dz)^2 dz$ and $c = 0$ is the only compatible solution
at the stationary state.
It is worth mentioning that when reaction rates are in the form of $k_u = k_u(u)$ and $k_v={\rm const.}$ as in Model I, 
$\mathscr{E}$ works as Lyapunov function commonly for $u$ and $p$~\cite{morita_stability_2010}.

\section{Analysis results for Model II}
In the main text, we investigated the pattern formation process for Model I. 
To examine model dependence of the results obtained there, 
in this section we analyze Model II and compare it with Model I. 
By performing the steady state analysis explained in the main text, 
we obtain results corresponding to Fig.~1{\bf c} and {\bf e}, presented in Fig.~\ref{fig:figS1}{\bf a} and {\bf b}, respectively. 
Unlike the case of Model I, the phase diagram for Model II does not have an upper bound of unstable and metastable regions about $T~(\equiv 1/RD)$. This is primarily because the nullcline of the reaction term $f(u,v) = 0$ has two branches that extend to infinitely large $v$,
and always crosses with the line $p_\infty = D_u u+D_v v$ at two points for any large value $T$ (see Fig.~\ref{fig:figS1}{\bf a}).
Note that we have set $k_0 = 0$ here, and drawing the full phase diagram for Model II requires an additional axis representing $k_0$.

Below we present the results of coarsening dynamics in 2D at point $(u, T)=(0.125, 1)$ (see the red cross symbol in Fig.~\ref{fig:figS1}{\bf b}). Fig.~\ref{fig:figS2}{\bf a} shows the temporal change of pattern (the density field of $s$), where droplet-like patterns form in the early time $(t=200)$ and coarsen over time.
Fig.~\ref{fig:figS3}{\bf a} shows the temporal change of the number of the droplets $n(t)$, which asymptotically approaches a power-law function with exponent $-0.60$ (red line). 
Thus, the growth rate is $0.60/d=0.30$ ($d=2$), which is coincident with the results for Model I (Fig.~2{\bf c} and 3{\bf d}).

Fig.~\ref{fig:figS3}{\bf b} shows the scaled structure factor, where we can see that the structure factors at different time are scaled for both small and large $q$ after the system reaches the coarsening regime where $n(t)$ shows a power-law decay. 
In Fig.~2{\bf d}, we have seen that the scalability for large $q$ in intermediate time regime does not hold for Model I. This difference is ascribed to 
whether the density inside the droplets is saturated to the stable fixed point $s_2$ (see Fig~1{\bf a}). 
For the current parameter choice for Model II,  
the values of the bistable steady solution $s_1$ and $s_2$ are not so far from that of the initial homogeneous state. The density inside the droplets easily saturates to the stable points, and as a result, the system shows a self-similar coarsening. 

Fig.~\ref{fig:figS2}{\bf b} shows the map of $p$ at the same spatial region and time as Fig.~\ref{fig:figS2}{\bf a}, where we can see a trend similar as Fig.~4b. This similarity, together with the growth exponent $0.30$, indicates that the evaporation-condensation mechanism also works in Model II. 
Moreover, in Fig.~\ref{fig:figS2}{\bf c}, one-to-one correspondence between $p$ and the droplet size by a power-law with exponent $-1/d$ is also confirmed as in Model I (see Fig.~4{\bf d} and {\bf g}). 
These results indicate that a surface-tension-like quantity generally exists in MCRDs without depending on details of the reaction terms, and plays a crucial role in the coarsening process of the patterns.

\section{Derivation of reduction equation}
Here we provide derivation of a reduction form Eq.~\ref{eq:reductionEq} by taking account of the slow dynamical mode associated with mass conservation and spatial inversion symmetry.

\subsection*{Basic property of MCRDs}
We consider the equations for MCRDs closed with $p = D_u u + D_v v$, $s = u+v$ (Eqs.~\ref{eq:pTime} and \ref{eq:conseved} in the main text).
Volume of the system $V = L^d$ ($d$ being the spatial dimension of the system) is sufficiently large and a periodic boundary condition is assumed.
\begin{align}
	s_0 = \frac{1}{V} \int s d{\bm r}
\end{align}
is the conserved quantity of the system.
As discussed in the main text $p(t,\bm r)$ becomes constant at the eventual stationary state $p = P(s_0)$, which depends on the conserved quantity $s_0$. For given $s_0$, $s$ at the stationary state is the solution of
\begin{align}
	\label{eq:stationaryS}
	-D_uD_v \nabla^2 s + (-D_u+D_v)f(P(s_0), s) = 0 ~.
\end{align}
We denote this stationary solution as $s(\bm r) = S({\bm r},s_0)$.
The stationary solution $(p,s) = \left(P(s_0), S({\bm r},s_0)\right)$ is parameterized by $s_0$ and constitutes a center manifold in the space of $(p,s)$. The derivative of the stationary solution with respect to $s_0$,
\begin{align}
	\label{eq:Szeromode}
 {\bm \eta}_s = \left(\frac{\partial P(s_0)}{\partial s_0}, \frac{\partial S(\bm r, s_0)}{\partial s_0} \right) ~,
\end{align}
is the tangential vector of the manifold, and is a zero-eigenmode of a liner operator $\mathcal{L}$, $\mathcal{L}{\bm \eta_s} = 0$, where $\mathcal{L}$ is 
derived from Eqs.~\ref{eq:pTime} and \ref{eq:conseved},
\begin{widetext}
\begin{align}
	\mathcal{L} =
	\begin{pmatrix}
		~~(D_u+D_v)\nabla^2 + (-D_u +D_v)\dfrac{\partial f}{\partial p} ~~~& -D_uD_v \nabla^2 +(-D_u+D_v)\dfrac{\partial f}{\partial s}~~ \\
		\nabla^2 & 0
	\end{pmatrix}~.
\end{align}
\end{widetext}
The conjugate vector of $\bm \eta_s$ 
is $\bm \eta^s = \left( 0, 1\right)$ for which the following relationship is shown.
\begin{align}
	\mathcal{L}^{\dagger} \bm \eta^s &= 0~, \\
	\langle \bm \eta^s , \bm \eta_s \rangle & = 1~.
\end{align}
Here, $\mathcal{L}^{\dagger}$ is the conjugate operator of $\mathcal{L}$, and
the scalar product between ${\bm a}(\bm r)$ and ${\bm b}(\bm r)$ is defined by
\begin{align}
\label{eq:InnerProduct}
\left\langle {\bm a},{\bm b}\right\rangle \equiv
\frac{1}{V}\int_V {\bm a}(\bm r)\cdot {\bm b}(\bm r) d \bm r,
\end{align}
All the relationships presented above are general consequences of mass conservation.

Note that there exist other zero-eigenmodes of $\mathcal{L}$, 
such as ${\bm \eta}_{x} = \left(0, \partial_{x} S(\bm r, s_0\right))$ 
that originates from the spatial translation invariance of the system
along $x$-axis. 
Similarly, ${\bm \eta}_{y} = \left(0, \partial_{y} S(\bm r, s_0)\right)$, etc., 
are zero-eigenmodes in multi-dimension.
They are irrelevant to the following analysis.

\subsection*{Dynamics on the center manifold}
We derive a single variable equation from Eqs.~\ref{eq:pTime} and \ref{eq:conseved} to describe the coarsening process of droplets based on the center manifold projection reduction method~\cite{kuramoto2003chemical, guckenheimer2013nonlinear}. 
As $p$ and $s$ in late stage are supposed to be quasi-static, we put an ansatz that they are expressed in the following form:
\begin{align}
	p(t, \bm r) & = P(c) + \rho_p(\bm r, c)~, \label{eq:pDecomp}\\
	s(t, \bm r) & = S(\bm r, c) + \rho_s (\bm r, c). \label{eq:sDecomp}
\end{align}
In each equation, the first term describes the dynamics along the center manifold,
and the second is the correction that is orthogonal to it. 
Therefore, ${\bm \rho} \equiv \left( \rho_p, \rho_s\right)$ is constrained by $\langle \bm \eta^s, \bm \rho\rangle = 0$ and $\langle \bm \eta^s, \partial_t \bm \rho\rangle = 0$. 
$c$ is ``coarse-grained mass density'', defined as 
\begin{equation}
	c(t, \bm X) \equiv \frac{1}{\tilde{V}} \int_{\tilde{V}} s d{\bm r}~.
\end{equation}
Here, $\tilde{V}$ is a volume element centered at $\bm X$, taken to be larger than the interfacial width but smaller than a typical center-of-mass distance between droplets. ${\bm X}$ is a coordinate to represent corresponding large-scale spatial variation of $c$.
Then we employ the multi-scale variable method 
\cite{kuramoto2003chemical, guckenheimer2013nonlinear} in which
${\bm r}$ and ${\bm X}$ are treated as independent variables, and  
$\nabla$ is replaced by $\nabla + \epsilon \nabla_X$; 
$\nabla_X$ is defined by derivatives with respect to ${\bm X}$ and $\epsilon$ is a small parameter representing gradual spatial variation of $c$.
By substituting Eqs.~\ref{eq:pDecomp} and \ref{eq:sDecomp} into 
Eqs.~\ref{eq:pTime} and \ref{eq:conseved}, one obtains the following equations
\begin{widetext}
\begin{align}
\label{eq:rhoequation}
& \partial_t {\bm \rho} - \mathcal{L} {\bm \rho} = -({\partial_t} c) {\bm \eta_s}+{\bm g}~,\\
&{\bm g} =  
\begin{pmatrix}
 (D_u+D_v) \left( \epsilon^2\nabla_X^2 P + \left(2\epsilon \nabla \!\cdot\! \nabla_X+ \epsilon^2\nabla_X^2 \right)\rho_p\right)
 -D_uD_v (2\epsilon \nabla_X \cdot \nabla + \epsilon^2 \nabla_X^2)
 \left(S + \rho_s\right) + (-D_u+D_v)\mathcal{N} \\
 \epsilon^2 \nabla_X^2 P+
(2\epsilon \nabla_X \cdot \nabla + \epsilon^2 \nabla_X^2) \rho_p
\end{pmatrix},
\end{align}
\end{widetext}
where $\mathcal{N}$ is the non-linear part of $f$ defined by
\begin{align}
 f(P+\rho_p , S+\rho_s) =f(P , S) +\frac{\partial f}{\partial p} \rho_p 
 +\frac{\partial f}{\partial s} \rho_s + \mathcal{N}(\bm \rho)~.
\end{align}
By taking scalar product between $\bm \eta^s$ and Eq.~\ref{eq:rhoequation}, and using
the relations $\langle \bm \eta^s, \partial_t \bm \rho\rangle = 0$,
$\langle \bm \eta^s, \mathcal{L} \bm \rho \rangle = 
\langle \mathcal{L}^{\dagger}\bm \eta^s,  \bm \rho\rangle= 0$, and
$\langle \bm \eta^s, \nabla \bm a\rangle = 0$ for an arbitrary function $\bm a (\bm r)$,
we obtain
\begin{align}
\label{eq:pdSdt}
	\frac{\partial c}{\partial t}& = \epsilon^2 \nabla_X^2 P(c) 	
	+ \epsilon^2 \nabla_X^2 \mathcal{B}(c)~,\\
\label{eq:pdSdt2}
	\mathcal{B}(c) &\equiv \frac{1}{V} \int_V  \rho_p (\bm r,c)d \bm r~.
\end{align}
Eq.~\ref{eq:pdSdt} provides the mass transport equation in this system.
The first term is the lowest order estimation 
of $\partial c/\partial t$. 
For evaluating higher term, our aim below is to analyze the behavior of $\rho_{p}$.
We expand ${\bm \rho}$ by $\epsilon$ and
evaluate $\rho_p$ for higher orders.
\begin{align}
{\bm \rho} &= \epsilon {\bm \rho}^{(1)} + \epsilon^2 {\bm \rho}^{(2)} + \cdots~.
\end{align}
Here, each $\bm \rho^{(i)}$ is supposed to be orthogonal to $\bm \eta^s$, i.e., $\langle \bm \eta^s, \bm \rho^{(i)}\rangle = \langle \bm \eta^s, \partial_t \bm \rho^{(i)}\rangle = 0$. 
Note that $\partial_t {\bm \rho}= (\partial \bm \rho/\partial c) (\partial c/\partial t) = \mathcal{O}(\epsilon^3)$ 
and does not contribute to $\mathcal{B}$ up to the second-order analysis performed below.

At the first order, Eq.~\ref{eq:rhoequation} leads into
\begin{align}
	\label{eq:1storderRho}
 -\mathcal{L} \bm \rho^{(1)} = \bm g^{(1)} \equiv 
\begin{pmatrix}
 -2D_uD_v \nabla_X \cdot \nabla S({\bm r},c) \\
0
\end{pmatrix}.
\end{align}
The solution of the linear equation is given by
\begin{align}
   \label{eq:decomposeRho1}
	\bm \rho^{(1)} = a^{(1)} \bm \eta_s +  \bm \eta^{(1)}~,
\end{align}
where $\bm \eta_s$ is the homogeneous solution given by Eq.~\ref{eq:Szeromode}, 
while $\bm \eta^{(1)}$ is the particular solution for given $\bm g^{(1)}$ in Eq.~\ref{eq:1storderRho}.
$a^{(1)}$ is constant with respect to $\bm r$, and is determined to satisfy the orthogonality condition 
$\langle \bm \eta^s, \bm \rho^{(1)} \rangle = 0$.
The second line of Eq.~\ref{eq:1storderRho}, $\nabla^2 \rho^{(1)}_{p} = 0$ indicates that $\rho^{(1)}_{p}$ is independent of ${\bm r}$.
By taking into account the first line of Eq.~\ref{eq:1storderRho}, we find that the solution is ascribed to 
the homogeneous one in Eq.~\ref{eq:decomposeRho1}, i.e., $a^{(1)} \propto \rho^{(1)}_{p}$. 
Notice that $\bm \eta_s$ and $\bm \eta^{(1)}$ have opposite parity against spatial reflection 
${\bm r} \to -{\bm r}$, i.e.,
$\bm \eta_s (S(\bm r), \bm r) \to  \bm \eta_s (S(-\bm r), \bm r) $ and
$\bm \eta^{(1)} (S(\bm r),\bm r) \to  - \bm \eta^{(1)}(S(-\bm r),\bm r) $, 
because $\bm g^{(1)}$ consists of only the first order derivative about ${\bm r}$.
The inner product between the functions with opposite parity vanishes; therefore, 
$a^{(1)} = - \langle \bm \eta^s, \bm \eta^{(1)}\rangle = 0$. 
Thus, $\mathcal{B}(c)$ in Eq.~\ref{eq:pdSdt} is zero at this order. 
This conclusion is simply understood by the fact that the system is invariant against spatial reflection,
so that the equation should not include odd terms about spatial derivative $\epsilon \nabla_X$,  excluding odd orders of $\epsilon$.
We can also determine $\rho^{(1)}_s(\bm r, c)$ from Eq.~\ref{eq:1storderRho}
in the same way, although its explicit form is not necessary below.

At the second order of expansion, Eq.~\ref{eq:rhoequation} leads into
\begin{widetext}
\begin{align}
	-\mathcal{L} \bm \rho^{(2)} =
\begin{pmatrix}
\left(D_u+D_v-\partial_c P \right)\nabla_X^2P -D_uD_v \left( 2\nabla_X\cdot\nabla \rho^{(1)}_{s}+\nabla_X^2 S\right)
+\dfrac{-D_u+D_v}{2}\dfrac{\partial^2 f}{\partial^2 s^2} \left(\rho^{(1)}_{s}\right)^2
\\
\left(1-\partial_c S \right) \nabla_X^2 P
\end{pmatrix}
,
\end{align}
\end{widetext}
where $\partial_c P = \partial P(c)/\partial c$ and $\partial_c S = \partial S(\bm r,c)/\partial c $. 
The second line of the above equality provides an equation for $\rho^{(2)}_p$, given by
\begin{align}
	\nabla^2 \rho^{(2)}_{p}(\bm r,c) = \left[\partial_c S(\bm r,c)-1\right] \nabla_X^2 P(c)~. 
\end{align}
The solution of $\rho^{(2)}_{p}$ has the form $\rho^{(2)}_{p} = \alpha(\bm r,c) \nabla_X^2 P(c)$, 
where $\alpha(\bm r, c) $ is the function of $\bm r$ and $c$. 
By using $\displaystyle{\mathcal{A}(c) = \frac{1}{V}\int  \alpha(\bm r,c)  d\bm r}$,
$\mathcal{B}(S)$ at the forth order is then given by
\begin{align}
 \mathcal{B}(c) = \frac{1}{V}\int_V \rho_p \nabla_X^2 P(c)d\bm r  
= \epsilon^2 \mathcal{A}(c) \nabla_X^2P(c) +\mathcal{O}(\epsilon^4) .
\end{align}
The higher term of $\mathcal{B}(c)$ is $\mathcal{O}(\epsilon^4)$, compatible with system invariance 
against spatial reflection.

By substituting $\mathcal{B}(c)$ into Eq.~\ref{eq:pdSdt}, we reach the following equation for $c$:
\begin{equation}
\begin{split}
	\frac{\partial c}{\partial t} = 
	\epsilon^2 \nabla_X^2 P(c) 	
	+ \epsilon^4 \nabla_X^2 \left[  \mathcal{A}(c) \nabla_X^2 P(c)\right] + \mathcal{O}(\epsilon^6)~,
\end{split}
\end{equation}
which describes the mass transport process between droplets. The second term corresponds to the perturbative term $\rho_p$ at the lowest (second) order.
Numerical data show that $p$ exhibits gradual spatial variation, compatible to center-of-mass distance among the droplets (see Fig.~\ref{fig:fig4}b and \ref{fig:figS2}b).
Thus, at the scale of our interest, we may replace $\epsilon \nabla_X$ and $c$ by $\nabla$ and $s$, respectively, and write a reduction form for $p$ as follows.
\begin{equation}
	p[s] = P(s) + \mathcal{A}(s) \nabla^2 P(s) + \mathcal{O}(\nabla^4)~.
\end{equation}

\section{Functional profile of $P(s)$ for Model I}
\begin{figure}[h]
\centering
\includegraphics[width=8.4cm]{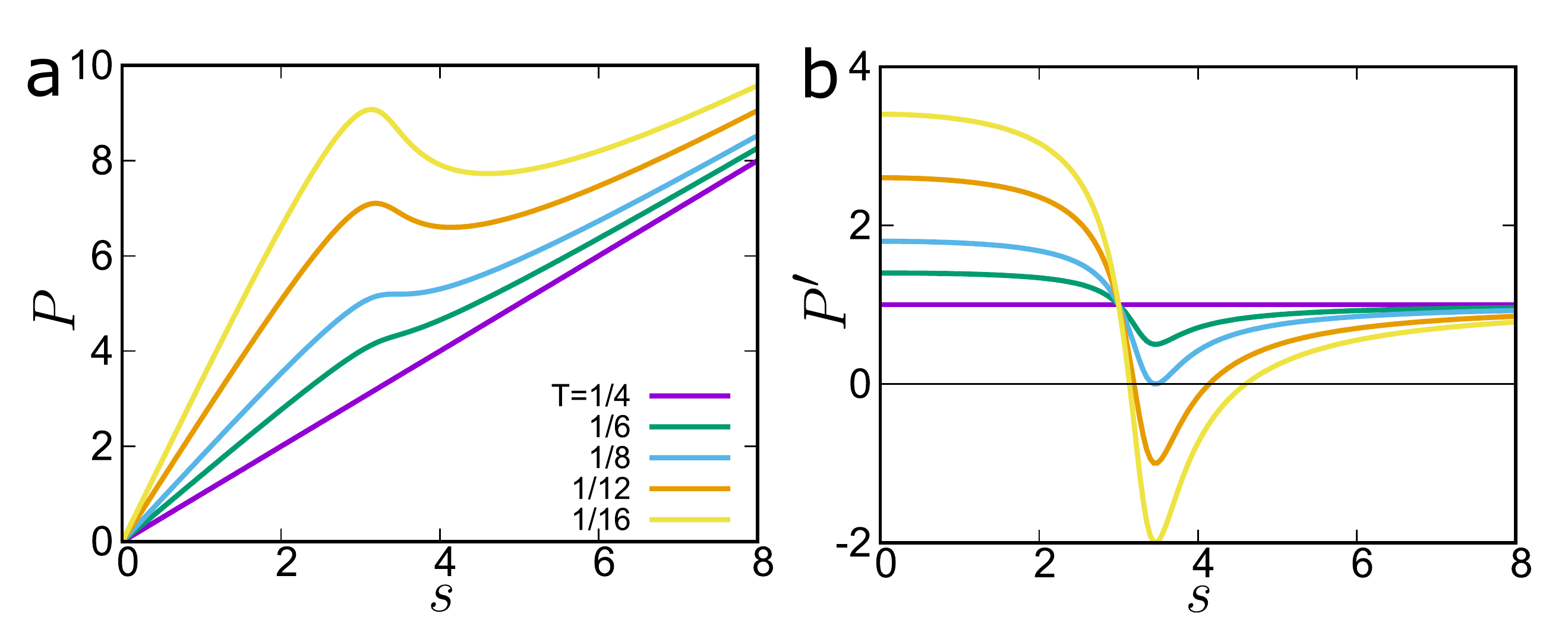}
\caption{{\bf Parameter dependence of the functional profile $P$} The functional profile $P$ for Model I for various $T$ and its first derivative are shown in panel {\bf a} and {\bf b}, respectively. Here we fix $R=1/4$ and vary $D$. In the figures, we can see that $P$ monotonically increases for $T>T_{\rm c}=1/8$ whereas for $T<T_{\rm c}$, the region $s$ exists where $P=P(s)$ has three solutions. 
}
\label{fig:figS4}
\end{figure}

$\mathcal{A}$ is the function of $s$ determined by 
steady stable solution $S$.
However, it is usually difficult to obtain their exact expression for given MCRDs.   
On the other side, $P(s)$ can be rather easily obtained since $P$ coincides with $p$ for a homogeneous steady solution $(u_0, v_0)$. In other words, $P$ satisfies $P=D_u u_0 + D_v v_0$ where 
\begin{eqnarray*}
f(u_0,v_0)&=0, \\
u_0+v_0&=s. 
\end{eqnarray*}
For Model I, $u_0$ is a solution for $u_0^3-su_0^2+(a+b)u_0-sb=0$. When the left-hand side of the cubic equation is monotonic for $u_0$, a unique (real) solution $u_0$ exists for any $s$, and thus we can analytically determine the specific form of $P$.  
In Fig.~\ref{fig:figS4}, we show the functional profiles $P$ for various parameter sets.
The functional profile of $P$ changes from a monotonically increasing function to non-monotonic one as dimensionless parameter $T$ crosses the critical value $T_c=1/8$.

%


\begin{thebibliography}{53}%
\makeatletter
\providecommand \@ifxundefined [1]{%
 \@ifx{#1\undefined}
}%
\providecommand \@ifnum [1]{%
 \ifnum #1\expandafter \@firstoftwo
 \else \expandafter \@secondoftwo
 \fi
}%
\providecommand \@ifx [1]{%
 \ifx #1\expandafter \@firstoftwo
 \else \expandafter \@secondoftwo
 \fi
}%
\providecommand \natexlab [1]{#1}%
\providecommand \enquote  [1]{``#1''}%
\providecommand \bibnamefont  [1]{#1}%
\providecommand \bibfnamefont [1]{#1}%
\providecommand \citenamefont [1]{#1}%
\providecommand \href@noop [0]{\@secondoftwo}%
\providecommand \href [0]{\begingroup \@sanitize@url \@href}%
\providecommand \@href[1]{\@@startlink{#1}\@@href}%
\providecommand \@@href[1]{\endgroup#1\@@endlink}%
\providecommand \@sanitize@url [0]{\catcode `\\12\catcode `\$12\catcode
  `\&12\catcode `\#12\catcode `\^12\catcode `\_12\catcode `\%12\relax}%
\providecommand \@@startlink[1]{}%
\providecommand \@@endlink[0]{}%
\providecommand \url  [0]{\begingroup\@sanitize@url \@url }%
\providecommand \@url [1]{\endgroup\@href {#1}{\urlprefix }}%
\providecommand \urlprefix  [0]{URL }%
\providecommand \Eprint [0]{\href }%
\providecommand \doibase [0]{http://dx.doi.org/}%
\providecommand \selectlanguage [0]{\@gobble}%
\providecommand \bibinfo  [0]{\@secondoftwo}%
\providecommand \bibfield  [0]{\@secondoftwo}%
\providecommand \translation [1]{[#1]}%
\providecommand \BibitemOpen [0]{}%
\providecommand \bibitemStop [0]{}%
\providecommand \bibitemNoStop [0]{.\EOS\space}%
\providecommand \EOS [0]{\spacefactor3000\relax}%
\providecommand \BibitemShut  [1]{\csname bibitem#1\endcsname}%
\let\auto@bib@innerbib\@empty
\bibitem [{\citenamefont {Cross}\ and\ \citenamefont
  {Hohenberg}(1993)}]{cross_pattern_1993}%
  \BibitemOpen
  \bibfield  {author} {\bibinfo {author} {\bibfnamefont {M.~C.}\ \bibnamefont
  {Cross}}\ and\ \bibinfo {author} {\bibfnamefont {P.~C.}\ \bibnamefont
  {Hohenberg}},\ }\href {\doibase 10.1103/RevModPhys.65.851} {\bibfield
  {journal} {\bibinfo  {journal} {Rev. Mod. Phys.}\ }\textbf {\bibinfo {volume}
  {65}},\ \bibinfo {pages} {851} (\bibinfo {year} {1993})}\BibitemShut
  {NoStop}%
\bibitem [{\citenamefont {Bressloff}\ \emph {et~al.}(2002)\citenamefont
  {Bressloff}, \citenamefont {Cowan}, \citenamefont {Golubitsky}, \citenamefont
  {Thomas},\ and\ \citenamefont {Wiener}}]{bressloff_what_2002}%
  \BibitemOpen
  \bibfield  {author} {\bibinfo {author} {\bibfnamefont {P.~C.}\ \bibnamefont
  {Bressloff}}, \bibinfo {author} {\bibfnamefont {J.~D.}\ \bibnamefont
  {Cowan}}, \bibinfo {author} {\bibfnamefont {M.}~\bibnamefont {Golubitsky}},
  \bibinfo {author} {\bibfnamefont {P.~J.}\ \bibnamefont {Thomas}}, \ and\
  \bibinfo {author} {\bibfnamefont {M.~C.}\ \bibnamefont {Wiener}},\ }\href
  {\doibase 10.1162/089976602317250861} {\bibfield  {journal} {\bibinfo
  {journal} {Neural Comput.}\ }\textbf {\bibinfo {volume} {14}},\ \bibinfo
  {pages} {473} (\bibinfo {year} {2002})}\BibitemShut {NoStop}%
\bibitem [{\citenamefont {Hoyle}(2006)}]{hoyle_pattern_2006}%
  \BibitemOpen
  \bibfield  {author} {\bibinfo {author} {\bibfnamefont {R.}~\bibnamefont
  {Hoyle}},\ }\href@noop {} {\emph {\bibinfo {title} {Pattern {Formation}: {An}
  {Introduction} to {Methods}}}}\ (\bibinfo  {publisher} {Cambridge University
  Press},\ \bibinfo {year} {2006})\BibitemShut {NoStop}%
\bibitem [{\citenamefont {Meron}(2016)}]{meron_pattern_2016}%
  \BibitemOpen
  \bibfield  {author} {\bibinfo {author} {\bibfnamefont {E.}~\bibnamefont
  {Meron}},\ }\href {\doibase 10.1016/j.mbs.2015.10.015} {\bibfield  {journal}
  {\bibinfo  {journal} {Math. Biosci.}\ }\textbf {\bibinfo {volume} {271}},\
  \bibinfo {pages} {1} (\bibinfo {year} {2016})}\BibitemShut {NoStop}%
\bibitem [{\citenamefont {Otsuji}\ \emph {et~al.}(2007)\citenamefont {Otsuji},
  \citenamefont {Ishihara}, \citenamefont {Co}, \citenamefont {Kaibuchi},
  \citenamefont {Mochizuki},\ and\ \citenamefont {Kuroda}}]{otsuji_mass_2007}%
  \BibitemOpen
  \bibfield  {author} {\bibinfo {author} {\bibfnamefont {M.}~\bibnamefont
  {Otsuji}}, \bibinfo {author} {\bibfnamefont {S.}~\bibnamefont {Ishihara}},
  \bibinfo {author} {\bibfnamefont {C.}~\bibnamefont {Co}}, \bibinfo {author}
  {\bibfnamefont {K.}~\bibnamefont {Kaibuchi}}, \bibinfo {author}
  {\bibfnamefont {A.}~\bibnamefont {Mochizuki}}, \ and\ \bibinfo {author}
  {\bibfnamefont {S.}~\bibnamefont {Kuroda}},\ }\href {\doibase
  10.1371/journal.pcbi.0030108} {\bibfield  {journal} {\bibinfo  {journal}
  {PLoS Comput. Biol}\ }\textbf {\bibinfo {volume} {3}},\ \bibinfo {pages}
  {e108} (\bibinfo {year} {2007})}\BibitemShut {NoStop}%
\bibitem [{\citenamefont {Ishihara}\ \emph {et~al.}(2007)\citenamefont
  {Ishihara}, \citenamefont {Otsuji},\ and\ \citenamefont
  {Mochizuki}}]{ishihara_transient_2007}%
  \BibitemOpen
  \bibfield  {author} {\bibinfo {author} {\bibfnamefont {S.}~\bibnamefont
  {Ishihara}}, \bibinfo {author} {\bibfnamefont {M.}~\bibnamefont {Otsuji}}, \
  and\ \bibinfo {author} {\bibfnamefont {A.}~\bibnamefont {Mochizuki}},\ }\href
  {\doibase 10.1103/PhysRevE.75.015203} {\bibfield  {journal} {\bibinfo
  {journal} {Phys. Rev. E}\ }\textbf {\bibinfo {volume} {75}},\ \bibinfo
  {pages} {015203} (\bibinfo {year} {2007})}\BibitemShut {NoStop}%
\bibitem [{\citenamefont {Mori}\ \emph {et~al.}(2008)\citenamefont {Mori},
  \citenamefont {Jilkine},\ and\ \citenamefont
  {Edelstein-Keshet}}]{mori_wave-pinning_2008}%
  \BibitemOpen
  \bibfield  {author} {\bibinfo {author} {\bibfnamefont {Y.}~\bibnamefont
  {Mori}}, \bibinfo {author} {\bibfnamefont {A.}~\bibnamefont {Jilkine}}, \
  and\ \bibinfo {author} {\bibfnamefont {L.}~\bibnamefont {Edelstein-Keshet}},\
  }\href {\doibase 10.1529/biophysj.107.120824} {\bibfield  {journal} {\bibinfo
   {journal} {Biophys. J.}\ }\textbf {\bibinfo {volume} {94}},\ \bibinfo
  {pages} {3684} (\bibinfo {year} {2008})}\BibitemShut {NoStop}%
\bibitem [{\citenamefont {Goryachev}\ and\ \citenamefont
  {Pokhilko}(2008)}]{goryachev_dynamics_2008}%
  \BibitemOpen
  \bibfield  {author} {\bibinfo {author} {\bibfnamefont {A.~B.}\ \bibnamefont
  {Goryachev}}\ and\ \bibinfo {author} {\bibfnamefont {A.~V.}\ \bibnamefont
  {Pokhilko}},\ }\href {\doibase 10.1016/j.febslet.2008.03.029} {\bibfield
  {journal} {\bibinfo  {journal} {FEBS Lett.}\ }\textbf {\bibinfo {volume}
  {582}},\ \bibinfo {pages} {1437} (\bibinfo {year} {2008})}\BibitemShut
  {NoStop}%
\bibitem [{\citenamefont {Rubinstein}\ \emph {et~al.}(2012)\citenamefont
  {Rubinstein}, \citenamefont {Slaughter},\ and\ \citenamefont
  {Li}}]{rubinstein2012weakly}%
  \BibitemOpen
  \bibfield  {author} {\bibinfo {author} {\bibfnamefont {B.}~\bibnamefont
  {Rubinstein}}, \bibinfo {author} {\bibfnamefont {B.~D.}\ \bibnamefont
  {Slaughter}}, \ and\ \bibinfo {author} {\bibfnamefont {R.}~\bibnamefont
  {Li}},\ }\href@noop {} {\bibfield  {journal} {\bibinfo  {journal} {Phys.
  Biol.}\ }\textbf {\bibinfo {volume} {9}},\ \bibinfo {pages} {045006}
  (\bibinfo {year} {2012})}\BibitemShut {NoStop}%
\bibitem [{\citenamefont {Holmes}\ and\ \citenamefont
  {Edelstein-Keshet}(2016)}]{holmes_analysis_2016}%
  \BibitemOpen
  \bibfield  {author} {\bibinfo {author} {\bibfnamefont {W.~R.}\ \bibnamefont
  {Holmes}}\ and\ \bibinfo {author} {\bibfnamefont {L.}~\bibnamefont
  {Edelstein-Keshet}},\ }\href {\doibase 10.1088/1478-3975/13/4/046001}
  {\bibfield  {journal} {\bibinfo  {journal} {Phys. Biol.}\ }\textbf {\bibinfo
  {volume} {13}},\ \bibinfo {pages} {046001} (\bibinfo {year}
  {2016})}\BibitemShut {NoStop}%
\bibitem [{\citenamefont {Chiou}\ \emph {et~al.}(2018)\citenamefont {Chiou},
  \citenamefont {Ramirez}, \citenamefont {Elston}, \citenamefont {Witelski},
  \citenamefont {Schaeffer},\ and\ \citenamefont
  {Lew}}]{chiou_principles_2018}%
  \BibitemOpen
  \bibfield  {author} {\bibinfo {author} {\bibfnamefont {J.-G.}\ \bibnamefont
  {Chiou}}, \bibinfo {author} {\bibfnamefont {S.~A.}\ \bibnamefont {Ramirez}},
  \bibinfo {author} {\bibfnamefont {T.~C.}\ \bibnamefont {Elston}}, \bibinfo
  {author} {\bibfnamefont {T.~P.}\ \bibnamefont {Witelski}}, \bibinfo {author}
  {\bibfnamefont {D.~G.}\ \bibnamefont {Schaeffer}}, \ and\ \bibinfo {author}
  {\bibfnamefont {D.~J.}\ \bibnamefont {Lew}},\ }\href {\doibase
  10.1371/journal.pcbi.1006095} {\bibfield  {journal} {\bibinfo  {journal}
  {PLoS Comput. Biol}\ }\textbf {\bibinfo {volume} {14}},\ \bibinfo {pages}
  {e1006095} (\bibinfo {year} {2018})}\BibitemShut {NoStop}%
\bibitem [{\citenamefont {Jacobs}\ \emph {et~al.}(2019)\citenamefont {Jacobs},
  \citenamefont {Molenaar},\ and\ \citenamefont {Deinum}}]{jacobs_small_2019}%
  \BibitemOpen
  \bibfield  {author} {\bibinfo {author} {\bibfnamefont {B.}~\bibnamefont
  {Jacobs}}, \bibinfo {author} {\bibfnamefont {J.}~\bibnamefont {Molenaar}}, \
  and\ \bibinfo {author} {\bibfnamefont {E.~E.}\ \bibnamefont {Deinum}},\
  }\href {\doibase 10.1371/journal.pone.0213188} {\bibfield  {journal}
  {\bibinfo  {journal} {PLoS One}\ }\textbf {\bibinfo {volume} {14}},\ \bibinfo
  {pages} {e0213188} (\bibinfo {year} {2019})}\BibitemShut {NoStop}%
\bibitem [{\citenamefont {Brauns}\ \emph {et~al.}()\citenamefont {Brauns},
  \citenamefont {Halatek},\ and\ \citenamefont {Frey}}]{brauns2018phase}%
  \BibitemOpen
  \bibfield  {author} {\bibinfo {author} {\bibfnamefont {F.}~\bibnamefont
  {Brauns}}, \bibinfo {author} {\bibfnamefont {J.}~\bibnamefont {Halatek}}, \
  and\ \bibinfo {author} {\bibfnamefont {E.}~\bibnamefont {Frey}},\ }\href@noop
  {} {\bibinfo  {journal} {Phy. Rev. X}, \ \bibinfo {pages}
  {accepted}}\BibitemShut {NoStop}%
\bibitem [{\citenamefont {Brauns}\ \emph {et~al.}(2020)\citenamefont {Brauns},
  \citenamefont {Weyer}, \citenamefont {Halatek}, \citenamefont {Yoon},\ and\
  \citenamefont {Frey}}]{brauns2020wavelength}%
  \BibitemOpen
\bibfield  {journal} {  }\bibfield  {author} {\bibinfo {author} {\bibfnamefont
  {F.}~\bibnamefont {Brauns}}, \bibinfo {author} {\bibfnamefont
  {H.}~\bibnamefont {Weyer}}, \bibinfo {author} {\bibfnamefont
  {J.}~\bibnamefont {Halatek}}, \bibinfo {author} {\bibfnamefont
  {J.}~\bibnamefont {Yoon}}, \ and\ \bibinfo {author} {\bibfnamefont
  {E.}~\bibnamefont {Frey}},\ }\href@noop {} {\bibfield  {journal} {\bibinfo
  {journal} {arXiv:2005.01495v2 [nlin.PS]}\ } (\bibinfo {year}
  {2020})}\BibitemShut {NoStop}%
\bibitem [{\citenamefont {Goryachev}\ and\ \citenamefont
  {Leda}(2020)}]{goryachev2020compete}%
  \BibitemOpen
  \bibfield  {author} {\bibinfo {author} {\bibfnamefont {A.~B.}\ \bibnamefont
  {Goryachev}}\ and\ \bibinfo {author} {\bibfnamefont {M.}~\bibnamefont
  {Leda}},\ }\href@noop {} {\bibfield  {journal} {\bibinfo  {journal} {Cells}\
  }\textbf {\bibinfo {volume} {9}},\ \bibinfo {pages} {2011} (\bibinfo {year}
  {2020})}\BibitemShut {NoStop}%
\bibitem [{\citenamefont {Othmer}\ and\ \citenamefont
  {Pate}(1980)}]{othmer_scale-invariance_1980}%
  \BibitemOpen
  \bibfield  {author} {\bibinfo {author} {\bibfnamefont {H.~G.}\ \bibnamefont
  {Othmer}}\ and\ \bibinfo {author} {\bibfnamefont {E.}~\bibnamefont {Pate}},\
  }\href {\doibase 10.1073/pnas.77.7.4180} {\bibfield  {journal} {\bibinfo
  {journal} {Proc. Natl. Acad. Sci. USA}\ }\textbf {\bibinfo {volume} {77}},\
  \bibinfo {pages} {4180} (\bibinfo {year} {1980})}\BibitemShut {NoStop}%
\bibitem [{\citenamefont {Hunding}\ and\ \citenamefont
  {Graae~S{\o}rensen}(1988)}]{hunding_size_1988}%
  \BibitemOpen
  \bibfield  {author} {\bibinfo {author} {\bibfnamefont {A.}~\bibnamefont
  {Hunding}}\ and\ \bibinfo {author} {\bibfnamefont {P.}~\bibnamefont
  {Graae~S{\o}rensen}},\ }\href {\doibase 10.1007/BF00280170} {\bibfield
  {journal} {\bibinfo  {journal} {J. Theor. Biol.}\ }\textbf {\bibinfo {volume}
  {26}},\ \bibinfo {pages} {27} (\bibinfo {year} {1988})}\BibitemShut {NoStop}%
\bibitem [{\citenamefont {Ishihara}\ and\ \citenamefont
  {Kaneko}(2006)}]{ishihara_turing_2006}%
  \BibitemOpen
  \bibfield  {author} {\bibinfo {author} {\bibfnamefont {S.}~\bibnamefont
  {Ishihara}}\ and\ \bibinfo {author} {\bibfnamefont {K.}~\bibnamefont
  {Kaneko}},\ }\href {\doibase 10.1016/j.jtbi.2005.06.016} {\bibfield
  {journal} {\bibinfo  {journal} {J. Theor. Biol.}\ }\textbf {\bibinfo {volume}
  {238}},\ \bibinfo {pages} {683} (\bibinfo {year} {2006})}\BibitemShut
  {NoStop}%
\bibitem [{\citenamefont {Ben-Zvi}\ and\ \citenamefont
  {Barkai}(2010)}]{ben-zvi_scaling_2010}%
  \BibitemOpen
  \bibfield  {author} {\bibinfo {author} {\bibfnamefont {D.}~\bibnamefont
  {Ben-Zvi}}\ and\ \bibinfo {author} {\bibfnamefont {N.}~\bibnamefont
  {Barkai}},\ }\href {\doibase 10.1073/pnas.0912734107} {\bibfield  {journal}
  {\bibinfo  {journal} {Proc. Natl. Acad. Sci. USA}\ }\textbf {\bibinfo
  {volume} {107}},\ \bibinfo {pages} {6924} (\bibinfo {year}
  {2010})}\BibitemShut {NoStop}%
\bibitem [{\citenamefont {Wartlick}\ \emph {et~al.}(2011)\citenamefont
  {Wartlick}, \citenamefont {Mumcu}, \citenamefont {Kicheva}, \citenamefont
  {Bittig}, \citenamefont {Seum}, \citenamefont {J{\"u}licher},\ and\
  \citenamefont {Gonz\'{a}lez-Gait\'{a}n}}]{wartlick_dynamics_2011}%
  \BibitemOpen
  \bibfield  {author} {\bibinfo {author} {\bibfnamefont {O.}~\bibnamefont
  {Wartlick}}, \bibinfo {author} {\bibfnamefont {P.}~\bibnamefont {Mumcu}},
  \bibinfo {author} {\bibfnamefont {A.}~\bibnamefont {Kicheva}}, \bibinfo
  {author} {\bibfnamefont {T.}~\bibnamefont {Bittig}}, \bibinfo {author}
  {\bibfnamefont {C.}~\bibnamefont {Seum}}, \bibinfo {author} {\bibfnamefont
  {F.}~\bibnamefont {J{\"u}licher}}, \ and\ \bibinfo {author} {\bibfnamefont
  {M.}~\bibnamefont {Gonz\'{a}lez-Gait\'{a}n}},\ }\href {\doibase
  10.1126/science.1200037} {\bibfield  {journal} {\bibinfo  {journal}
  {Science}\ }\textbf {\bibinfo {volume} {331}},\ \bibinfo {pages} {1154}
  (\bibinfo {year} {2011})}\BibitemShut {NoStop}%
\bibitem [{\citenamefont {Umulis}\ and\ \citenamefont
  {Othmer}(2013)}]{umulis_mechanisms_2013}%
  \BibitemOpen
  \bibfield  {author} {\bibinfo {author} {\bibfnamefont {D.~M.}\ \bibnamefont
  {Umulis}}\ and\ \bibinfo {author} {\bibfnamefont {H.~G.}\ \bibnamefont
  {Othmer}},\ }\href {\doibase 10.1242/dev.100511} {\bibfield  {journal}
  {\bibinfo  {journal} {Development}\ }\textbf {\bibinfo {volume} {140}},\
  \bibinfo {pages} {4830} (\bibinfo {year} {2013})}\BibitemShut {NoStop}%
\bibitem [{\citenamefont {Onuki}(2002)}]{onuki2002phase}%
  \BibitemOpen
  \bibfield  {author} {\bibinfo {author} {\bibfnamefont {A.}~\bibnamefont
  {Onuki}},\ }\href@noop {} {\emph {\bibinfo {title} {Phase Transition
  Dynamics}}}\ (\bibinfo  {publisher} {Cambridge University Press},\ \bibinfo
  {year} {2002})\BibitemShut {NoStop}%
\bibitem [{\citenamefont {Mori}\ and\ \citenamefont
  {Kuramoto}(2013)}]{mori2013dissipative}%
  \BibitemOpen
  \bibfield  {author} {\bibinfo {author} {\bibfnamefont {H.}~\bibnamefont
  {Mori}}\ and\ \bibinfo {author} {\bibfnamefont {Y.}~\bibnamefont
  {Kuramoto}},\ }\href@noop {} {\emph {\bibinfo {title} {Dissipative structures
  and chaos}}}\ (\bibinfo  {publisher} {Springer Science \& Business Media},\
  \bibinfo {year} {2013})\BibitemShut {NoStop}%
\bibitem [{\citenamefont {Hohenberg}\ and\ \citenamefont
  {Halperin}(1977)}]{hohenberg_theory_1977}%
  \BibitemOpen
  \bibfield  {author} {\bibinfo {author} {\bibfnamefont {P.~C.}\ \bibnamefont
  {Hohenberg}}\ and\ \bibinfo {author} {\bibfnamefont {B.~I.}\ \bibnamefont
  {Halperin}},\ }\href {\doibase 10.1103/RevModPhys.49.435} {\bibfield
  {journal} {\bibinfo  {journal} {Rev. Mod. Phys.}\ }\textbf {\bibinfo {volume}
  {49}},\ \bibinfo {pages} {435} (\bibinfo {year} {1977})}\BibitemShut
  {NoStop}%
\bibitem [{\citenamefont {Khanna}\ \emph {et~al.}(2010)\citenamefont {Khanna},
  \citenamefont {Agnihotri}, \citenamefont {Vashishtha}, \citenamefont
  {Sharma}, \citenamefont {Jaiswal},\ and\ \citenamefont
  {Puri}}]{khanna_kinetics_2010}%
  \BibitemOpen
  \bibfield  {author} {\bibinfo {author} {\bibfnamefont {R.}~\bibnamefont
  {Khanna}}, \bibinfo {author} {\bibfnamefont {N.~K.}\ \bibnamefont
  {Agnihotri}}, \bibinfo {author} {\bibfnamefont {M.}~\bibnamefont
  {Vashishtha}}, \bibinfo {author} {\bibfnamefont {A.}~\bibnamefont {Sharma}},
  \bibinfo {author} {\bibfnamefont {P.~K.}\ \bibnamefont {Jaiswal}}, \ and\
  \bibinfo {author} {\bibfnamefont {S.}~\bibnamefont {Puri}},\ }\href {\doibase
  10.1103/PhysRevE.82.011601} {\bibfield  {journal} {\bibinfo  {journal} {Phys.
  Rev. E}\ }\textbf {\bibinfo {volume} {82}},\ \bibinfo {pages} {011601}
  (\bibinfo {year} {2010})}\BibitemShut {NoStop}%
\bibitem [{\citenamefont {Narayanam}\ \emph {et~al.}(2017)\citenamefont
  {Narayanam}, \citenamefont {Kumar}, \citenamefont {Puri},\ and\ \citenamefont
  {Khanna}}]{narayanam_coarsening_2017}%
  \BibitemOpen
  \bibfield  {author} {\bibinfo {author} {\bibfnamefont {C.}~\bibnamefont
  {Narayanam}}, \bibinfo {author} {\bibfnamefont {A.}~\bibnamefont {Kumar}},
  \bibinfo {author} {\bibfnamefont {S.}~\bibnamefont {Puri}}, \ and\ \bibinfo
  {author} {\bibfnamefont {R.}~\bibnamefont {Khanna}},\ }\href {\doibase
  10.1021/acs.langmuir.7b00752} {\bibfield  {journal} {\bibinfo  {journal}
  {Langmuir}\ }\textbf {\bibinfo {volume} {33}},\ \bibinfo {pages} {3341}
  (\bibinfo {year} {2017})}\BibitemShut {NoStop}%
\bibitem [{\citenamefont {Zhang}\ \emph {et~al.}(2017)\citenamefont {Zhang},
  \citenamefont {Yang}, \citenamefont {Yang},\ and\ \citenamefont
  {Wang}}]{zhang_morphological-evolution_2017}%
  \BibitemOpen
  \bibfield  {author} {\bibinfo {author} {\bibfnamefont {G.}~\bibnamefont
  {Zhang}}, \bibinfo {author} {\bibfnamefont {T.}~\bibnamefont {Yang}},
  \bibinfo {author} {\bibfnamefont {S.}~\bibnamefont {Yang}}, \ and\ \bibinfo
  {author} {\bibfnamefont {Y.}~\bibnamefont {Wang}},\ }\href {\doibase
  10.1103/PhysRevE.96.032501} {\bibfield  {journal} {\bibinfo  {journal} {Phys.
  Rev. E}\ }\textbf {\bibinfo {volume} {96}},\ \bibinfo {pages} {032501}
  (\bibinfo {year} {2017})}\BibitemShut {NoStop}%
\bibitem [{\citenamefont {Yeung}(1988)}]{yeung_scaling_1988}%
  \BibitemOpen
  \bibfield  {author} {\bibinfo {author} {\bibfnamefont {C.}~\bibnamefont
  {Yeung}},\ }\href {\doibase 10.1103/PhysRevLett.61.1135} {\bibfield
  {journal} {\bibinfo  {journal} {Phys. Rev. Lett.}\ }\textbf {\bibinfo
  {volume} {61}},\ \bibinfo {pages} {1135} (\bibinfo {year}
  {1988})}\BibitemShut {NoStop}%
\bibitem [{\citenamefont {Fratzl}\ and\ \citenamefont
  {Lebowitz}(1989)}]{fratzl_quenched_1989}%
  \BibitemOpen
  \bibfield  {author} {\bibinfo {author} {\bibfnamefont {P.}~\bibnamefont
  {Fratzl}}\ and\ \bibinfo {author} {\bibfnamefont {J.~L.}\ \bibnamefont
  {Lebowitz}},\ }\href@noop {} {\bibfield  {journal} {\bibinfo  {journal} {Acta
  metall}\ }\textbf {\bibinfo {volume} {37}},\ \bibinfo {pages} {3245}
  (\bibinfo {year} {1989})}\BibitemShut {NoStop}%
\bibitem [{\citenamefont {Koga}\ \emph {et~al.}(1993)\citenamefont {Koga},
  \citenamefont {Kawasaki}, \citenamefont {Takenaka},\ and\ \citenamefont
  {Hashimoto}}]{koga_late_1993}%
  \BibitemOpen
  \bibfield  {author} {\bibinfo {author} {\bibfnamefont {T.}~\bibnamefont
  {Koga}}, \bibinfo {author} {\bibfnamefont {K.}~\bibnamefont {Kawasaki}},
  \bibinfo {author} {\bibfnamefont {M.}~\bibnamefont {Takenaka}}, \ and\
  \bibinfo {author} {\bibfnamefont {T.}~\bibnamefont {Hashimoto}},\ }\href
  {\doibase 10.1016/0378-4371(93)90235-V} {\bibfield  {journal} {\bibinfo
  {journal} {Physica A}\ }\textbf {\bibinfo {volume} {198}},\ \bibinfo {pages}
  {473} (\bibinfo {year} {1993})}\BibitemShut {NoStop}%
\bibitem [{\citenamefont {Furukawa}(1989)}]{furukawa_multi-time_1989}%
  \BibitemOpen
  \bibfield  {author} {\bibinfo {author} {\bibfnamefont {H.}~\bibnamefont
  {Furukawa}},\ }\href {\doibase 10.1143/JPSJ.58.216} {\bibfield  {journal}
  {\bibinfo  {journal} {J, Phys. Soc. Jpn.}\ }\textbf {\bibinfo {volume}
  {58}},\ \bibinfo {pages} {216} (\bibinfo {year} {1989})}\BibitemShut
  {NoStop}%
\bibitem [{\citenamefont {Tomita}(1991)}]{tomita_preservation_1991}%
  \BibitemOpen
  \bibfield  {author} {\bibinfo {author} {\bibfnamefont {H.}~\bibnamefont
  {Tomita}},\ }\href {\doibase 10.1143/ptp/85.1.47} {\bibfield  {journal}
  {\bibinfo  {journal} {Prog. Theor. Exp. Phys.}\ }\textbf {\bibinfo {volume}
  {85}},\ \bibinfo {pages} {47} (\bibinfo {year} {1991})},\ \bibinfo {note}
  {publisher: Oxford Academic}\BibitemShut {NoStop}%
\bibitem [{\citenamefont {Biancalani}\ \emph {et~al.}(2017)\citenamefont
  {Biancalani}, \citenamefont {Jafarpour},\ and\ \citenamefont
  {Goldenfeld}}]{biancalani_giant_2017}%
  \BibitemOpen
  \bibfield  {author} {\bibinfo {author} {\bibfnamefont {T.}~\bibnamefont
  {Biancalani}}, \bibinfo {author} {\bibfnamefont {F.}~\bibnamefont
  {Jafarpour}}, \ and\ \bibinfo {author} {\bibfnamefont {N.}~\bibnamefont
  {Goldenfeld}},\ }\href {\doibase 10.1103/PhysRevLett.118.018101} {\bibfield
  {journal} {\bibinfo  {journal} {Phys. Rev. Lett.}\ }\textbf {\bibinfo
  {volume} {118}},\ \bibinfo {pages} {018101} (\bibinfo {year}
  {2017})}\BibitemShut {NoStop}%
\bibitem [{\citenamefont {Kalikmanov}(2013)}]{kalikmanov_classical_2013}%
  \BibitemOpen
  \bibfield  {author} {\bibinfo {author} {\bibfnamefont {V.~I.}\ \bibnamefont
  {Kalikmanov}},\ }in\ \href {\doibase 10.1007/978-90-481-3643-8_3} {\emph
  {\bibinfo {booktitle} {Nucleation {Theory}}}},\ \bibinfo {series and number}
  {Lecture {Notes} in {Physics}}\ (\bibinfo  {publisher} {Springer
  Netherlands},\ \bibinfo {address} {Dordrecht},\ \bibinfo {year}
  {2013})\BibitemShut {NoStop}%
\bibitem [{\citenamefont {Butler}\ and\ \citenamefont
  {Goldenfeld}(2009)}]{butler_robust_2009}%
  \BibitemOpen
  \bibfield  {author} {\bibinfo {author} {\bibfnamefont {T.}~\bibnamefont
  {Butler}}\ and\ \bibinfo {author} {\bibfnamefont {N.}~\bibnamefont
  {Goldenfeld}},\ }\href {\doibase 10.1103/PhysRevE.80.030902} {\bibfield
  {journal} {\bibinfo  {journal} {Phys. Rev. E}\ }\textbf {\bibinfo {volume}
  {80}},\ \bibinfo {pages} {030902} (\bibinfo {year} {2009})}\BibitemShut
  {NoStop}%
\bibitem [{\citenamefont {Biancalani}\ \emph {et~al.}(2010)\citenamefont
  {Biancalani}, \citenamefont {Fanelli},\ and\ \citenamefont
  {Di~Patti}}]{biancalani_stochastic_2010}%
  \BibitemOpen
  \bibfield  {author} {\bibinfo {author} {\bibfnamefont {T.}~\bibnamefont
  {Biancalani}}, \bibinfo {author} {\bibfnamefont {D.}~\bibnamefont {Fanelli}},
  \ and\ \bibinfo {author} {\bibfnamefont {F.}~\bibnamefont {Di~Patti}},\
  }\href {\doibase 10.1103/PhysRevE.81.046215} {\bibfield  {journal} {\bibinfo
  {journal} {Phys. Rev. E}\ }\textbf {\bibinfo {volume} {81}},\ \bibinfo
  {pages} {046215} (\bibinfo {year} {2010})}\BibitemShut {NoStop}%
\bibitem [{\citenamefont {Bonachela}\ \emph {et~al.}(2012)\citenamefont
  {Bonachela}, \citenamefont {Mu{\~ n}oz},\ and\ \citenamefont
  {Levin}}]{bonachela_patchiness_2012}%
  \BibitemOpen
  \bibfield  {author} {\bibinfo {author} {\bibfnamefont {J.~A.}\ \bibnamefont
  {Bonachela}}, \bibinfo {author} {\bibfnamefont {M.~A.}\ \bibnamefont {Mu{\~
  n}oz}}, \ and\ \bibinfo {author} {\bibfnamefont {S.~A.}\ \bibnamefont
  {Levin}},\ }\href {\doibase 10.1007/s10955-012-0506-x} {\bibfield  {journal}
  {\bibinfo  {journal} {J. Stat. Mech.}\ }\textbf {\bibinfo {volume} {148}},\
  \bibinfo {pages} {724} (\bibinfo {year} {2012})}\BibitemShut {NoStop}%
\bibitem [{\citenamefont {Butler}\ and\ \citenamefont
  {Goldenfeld}(2011)}]{butler_fluctuation-driven_2011}%
  \BibitemOpen
  \bibfield  {author} {\bibinfo {author} {\bibfnamefont {T.}~\bibnamefont
  {Butler}}\ and\ \bibinfo {author} {\bibfnamefont {N.}~\bibnamefont
  {Goldenfeld}},\ }\href {\doibase 10.1103/PhysRevE.84.011112} {\bibfield
  {journal} {\bibinfo  {journal} {Phys. Rev. E}\ }\textbf {\bibinfo {volume}
  {84}},\ \bibinfo {pages} {011112} (\bibinfo {year} {2011})}\BibitemShut
  {NoStop}%
\bibitem [{\citenamefont {Karig}\ \emph {et~al.}(2018)\citenamefont {Karig},
  \citenamefont {Martini}, \citenamefont {Lu}, \citenamefont {DeLateur},
  \citenamefont {Goldenfeld},\ and\ \citenamefont
  {Weiss}}]{karig_stochastic_2018}%
  \BibitemOpen
  \bibfield  {author} {\bibinfo {author} {\bibfnamefont {D.}~\bibnamefont
  {Karig}}, \bibinfo {author} {\bibfnamefont {K.~M.}\ \bibnamefont {Martini}},
  \bibinfo {author} {\bibfnamefont {T.}~\bibnamefont {Lu}}, \bibinfo {author}
  {\bibfnamefont {N.~A.}\ \bibnamefont {DeLateur}}, \bibinfo {author}
  {\bibfnamefont {N.}~\bibnamefont {Goldenfeld}}, \ and\ \bibinfo {author}
  {\bibfnamefont {R.}~\bibnamefont {Weiss}},\ }\href@noop {} {\bibfield
  {journal} {\bibinfo  {journal} {Proc. Natl. Acad. Sci. USA}\ }\textbf
  {\bibinfo {volume} {115}} (\bibinfo {year} {2018})}\BibitemShut {NoStop}%
\bibitem [{\citenamefont {Solon}\ \emph {et~al.}(2018)\citenamefont {Solon},
  \citenamefont {Stenhammar}, \citenamefont {Cates}, \citenamefont {Kafri},\
  and\ \citenamefont {Tailleur}}]{solon_generalized_2018}%
  \BibitemOpen
  \bibfield  {author} {\bibinfo {author} {\bibfnamefont {A.~P.}\ \bibnamefont
  {Solon}}, \bibinfo {author} {\bibfnamefont {J.}~\bibnamefont {Stenhammar}},
  \bibinfo {author} {\bibfnamefont {M.~E.}\ \bibnamefont {Cates}}, \bibinfo
  {author} {\bibfnamefont {Y.}~\bibnamefont {Kafri}}, \ and\ \bibinfo {author}
  {\bibfnamefont {J.}~\bibnamefont {Tailleur}},\ }\href {\doibase
  10.1088/1367-2630/aaccdd} {\bibfield  {journal} {\bibinfo  {journal} {New J.
  Phys.}\ }\textbf {\bibinfo {volume} {20}},\ \bibinfo {pages} {075001}
  (\bibinfo {year} {2018})}\BibitemShut {NoStop}%
\bibitem [{\citenamefont {Chen}\ \emph {et~al.}(2020)\citenamefont {Chen},
  \citenamefont {Wu}, \citenamefont {Wu},\ and\ \citenamefont
  {Zhang}}]{chen_phase_2020}%
  \BibitemOpen
  \bibfield  {author} {\bibinfo {author} {\bibfnamefont {X.}~\bibnamefont
  {Chen}}, \bibinfo {author} {\bibfnamefont {X.}~\bibnamefont {Wu}}, \bibinfo
  {author} {\bibfnamefont {H.}~\bibnamefont {Wu}}, \ and\ \bibinfo {author}
  {\bibfnamefont {M.}~\bibnamefont {Zhang}},\ }\href@noop {} {\bibfield
  {journal} {\bibinfo  {journal} {Nature Neuroscience}\ ,\ \bibinfo {pages}
  {1}} (\bibinfo {year} {2020})},\ \bibinfo {note} {publisher: Nature
  Publishing Group}\BibitemShut {NoStop}%
\bibitem [{\citenamefont {Hansen}\ \emph {et~al.}(2019)\citenamefont {Hansen},
  \citenamefont {Huang}, \citenamefont {Lee}, \citenamefont {Bieling},
  \citenamefont {Christensen},\ and\ \citenamefont
  {Groves}}]{hansen_stochastic_2019}%
  \BibitemOpen
  \bibfield  {author} {\bibinfo {author} {\bibfnamefont {S.~D.}\ \bibnamefont
  {Hansen}}, \bibinfo {author} {\bibfnamefont {W.~Y.~C.}\ \bibnamefont
  {Huang}}, \bibinfo {author} {\bibfnamefont {Y.~K.}\ \bibnamefont {Lee}},
  \bibinfo {author} {\bibfnamefont {P.}~\bibnamefont {Bieling}}, \bibinfo
  {author} {\bibfnamefont {S.~M.}\ \bibnamefont {Christensen}}, \ and\ \bibinfo
  {author} {\bibfnamefont {J.~T.}\ \bibnamefont {Groves}},\ }\href {\doibase
  10.1073/pnas.1901744116} {\bibfield  {journal} {\bibinfo  {journal} {Proc.
  Natl. Acad. Sci. USA}\ }\textbf {\bibinfo {volume} {116}},\ \bibinfo {pages}
  {15013} (\bibinfo {year} {2019})}\BibitemShut {NoStop}%
\bibitem [{\citenamefont {Frey}\ \emph {et~al.}(2018)\citenamefont {Frey},
  \citenamefont {Halatek}, \citenamefont {Kretschmer},\ and\ \citenamefont
  {Schwille}}]{frey_protein_2018}%
  \BibitemOpen
  \bibfield  {author} {\bibinfo {author} {\bibfnamefont {E.}~\bibnamefont
  {Frey}}, \bibinfo {author} {\bibfnamefont {J.}~\bibnamefont {Halatek}},
  \bibinfo {author} {\bibfnamefont {S.}~\bibnamefont {Kretschmer}}, \ and\
  \bibinfo {author} {\bibfnamefont {P.}~\bibnamefont {Schwille}},\ }in\ \href
  {\doibase 10.1007/978-3-030-00630-3_10} {\emph {\bibinfo {booktitle} {Physics
  of {Biological} {Membranes}}}}\ (\bibinfo  {publisher} {Springer
  International Publishing},\ \bibinfo {address} {Cham},\ \bibinfo {year}
  {2018})\ pp.\ \bibinfo {pages} {229--260}\BibitemShut {NoStop}%
\bibitem [{\citenamefont {Halatek}\ and\ \citenamefont
  {Frey}(2018)}]{halatek_rethinking_2018}%
  \BibitemOpen
  \bibfield  {author} {\bibinfo {author} {\bibfnamefont {J.}~\bibnamefont
  {Halatek}}\ and\ \bibinfo {author} {\bibfnamefont {E.}~\bibnamefont {Frey}},\
  }\href {\doibase 10.1038/s41567-017-0040-5} {\bibfield  {journal} {\bibinfo
  {journal} {Nat. Phys.}\ }\textbf {\bibinfo {volume} {14}},\ \bibinfo {pages}
  {507} (\bibinfo {year} {2018})}\BibitemShut {NoStop}%
\bibitem [{\citenamefont {Hyman}\ \emph {et~al.}(2014)\citenamefont {Hyman},
  \citenamefont {Weber},\ and\ \citenamefont
  {J{\"{u}}licher}}]{hyman2014liquid}%
  \BibitemOpen
  \bibfield  {author} {\bibinfo {author} {\bibfnamefont {A.~A.}\ \bibnamefont
  {Hyman}}, \bibinfo {author} {\bibfnamefont {C.~A.}\ \bibnamefont {Weber}}, \
  and\ \bibinfo {author} {\bibfnamefont {F.}~\bibnamefont {J{\"{u}}licher}},\
  }\href@noop {} {\bibfield  {journal} {\bibinfo  {journal} {Annu. Rev. Cell
  Dev. Biol.}\ }\textbf {\bibinfo {volume} {30}},\ \bibinfo {pages} {39}
  (\bibinfo {year} {2014})}\BibitemShut {NoStop}%
\bibitem [{\citenamefont {Brangwynne}\ \emph {et~al.}(2015)\citenamefont
  {Brangwynne}, \citenamefont {Tompa},\ and\ \citenamefont
  {Pappu}}]{brangwynne2015polymer}%
  \BibitemOpen
  \bibfield  {author} {\bibinfo {author} {\bibfnamefont {C.~P.}\ \bibnamefont
  {Brangwynne}}, \bibinfo {author} {\bibfnamefont {P.}~\bibnamefont {Tompa}}, \
  and\ \bibinfo {author} {\bibfnamefont {R.~V.}\ \bibnamefont {Pappu}},\
  }\href@noop {} {\bibfield  {journal} {\bibinfo  {journal} {Nat. Phys.}\
  }\textbf {\bibinfo {volume} {11}},\ \bibinfo {pages} {899} (\bibinfo {year}
  {2015})}\BibitemShut {NoStop}%
\bibitem [{\citenamefont {Boeynaems}\ \emph {et~al.}(2018)\citenamefont
  {Boeynaems}, \citenamefont {Alberti}, \citenamefont {Fawzi}, \citenamefont
  {Mittag}, \citenamefont {Polymenidou}, \citenamefont {Rousseau},
  \citenamefont {Schymkowitz}, \citenamefont {Shorter}, \citenamefont
  {Wolozin}, \citenamefont {Van Den~Bosch} \emph
  {et~al.}}]{boeynaems2018protein}%
  \BibitemOpen
  \bibfield  {author} {\bibinfo {author} {\bibfnamefont {S.}~\bibnamefont
  {Boeynaems}}, \bibinfo {author} {\bibfnamefont {S.}~\bibnamefont {Alberti}},
  \bibinfo {author} {\bibfnamefont {N.~L.}\ \bibnamefont {Fawzi}}, \bibinfo
  {author} {\bibfnamefont {T.}~\bibnamefont {Mittag}}, \bibinfo {author}
  {\bibfnamefont {M.}~\bibnamefont {Polymenidou}}, \bibinfo {author}
  {\bibfnamefont {F.}~\bibnamefont {Rousseau}}, \bibinfo {author}
  {\bibfnamefont {J.}~\bibnamefont {Schymkowitz}}, \bibinfo {author}
  {\bibfnamefont {J.}~\bibnamefont {Shorter}}, \bibinfo {author} {\bibfnamefont
  {B.}~\bibnamefont {Wolozin}}, \bibinfo {author} {\bibfnamefont
  {L.}~\bibnamefont {Van Den~Bosch}},  \emph {et~al.},\ }\href@noop {}
  {\bibfield  {journal} {\bibinfo  {journal} {Trends Cell Biol.}\ }\textbf
  {\bibinfo {volume} {28}},\ \bibinfo {pages} {420} (\bibinfo {year}
  {2018})}\BibitemShut {NoStop}%
\bibitem [{\citenamefont {Berry}\ \emph {et~al.}(2018)\citenamefont {Berry},
  \citenamefont {Brangwynne},\ and\ \citenamefont
  {Haataja}}]{berry2018physical}%
  \BibitemOpen
  \bibfield  {author} {\bibinfo {author} {\bibfnamefont {J.}~\bibnamefont
  {Berry}}, \bibinfo {author} {\bibfnamefont {C.~P.}\ \bibnamefont
  {Brangwynne}}, \ and\ \bibinfo {author} {\bibfnamefont {M.}~\bibnamefont
  {Haataja}},\ }\href@noop {} {\bibfield  {journal} {\bibinfo  {journal} {Rep.
  Prog. Phys.}\ }\textbf {\bibinfo {volume} {81}},\ \bibinfo {pages} {046601}
  (\bibinfo {year} {2018})}\BibitemShut {NoStop}%
\bibitem [{\citenamefont {Cates}\ and\ \citenamefont
  {Tailleur}(2015)}]{cates2015motility}%
  \BibitemOpen
  \bibfield  {author} {\bibinfo {author} {\bibfnamefont {M.~E.}\ \bibnamefont
  {Cates}}\ and\ \bibinfo {author} {\bibfnamefont {J.}~\bibnamefont
  {Tailleur}},\ }\href@noop {} {\bibfield  {journal} {\bibinfo  {journal}
  {Annu. Rev. Condens. Matter Phys.}\ }\textbf {\bibinfo {volume} {6}},\
  \bibinfo {pages} {219} (\bibinfo {year} {2015})}\BibitemShut {NoStop}%
\bibitem [{\citenamefont {Gompper}\ \emph {et~al.}(2020)\citenamefont
  {Gompper}, \citenamefont {Winkler}, \citenamefont {Speck}, \citenamefont
  {Solon}, \citenamefont {Nardini}, \citenamefont {Peruani}, \citenamefont
  {L{\"o}wen}, \citenamefont {Golestanian}, \citenamefont {Kaupp},
  \citenamefont {Alvarez} \emph {et~al.}}]{gompper20202020}%
  \BibitemOpen
  \bibfield  {author} {\bibinfo {author} {\bibfnamefont {G.}~\bibnamefont
  {Gompper}}, \bibinfo {author} {\bibfnamefont {R.~G.}\ \bibnamefont
  {Winkler}}, \bibinfo {author} {\bibfnamefont {T.}~\bibnamefont {Speck}},
  \bibinfo {author} {\bibfnamefont {A.}~\bibnamefont {Solon}}, \bibinfo
  {author} {\bibfnamefont {C.}~\bibnamefont {Nardini}}, \bibinfo {author}
  {\bibfnamefont {F.}~\bibnamefont {Peruani}}, \bibinfo {author} {\bibfnamefont
  {H.}~\bibnamefont {L{\"o}wen}}, \bibinfo {author} {\bibfnamefont
  {R.}~\bibnamefont {Golestanian}}, \bibinfo {author} {\bibfnamefont {U.~B.}\
  \bibnamefont {Kaupp}}, \bibinfo {author} {\bibfnamefont {L.}~\bibnamefont
  {Alvarez}},  \emph {et~al.},\ }\href@noop {} {\bibfield  {journal} {\bibinfo
  {journal} {J. Phys. Condens. Matter}\ }\textbf {\bibinfo {volume} {32}},\
  \bibinfo {pages} {193001} (\bibinfo {year} {2020})}\BibitemShut {NoStop}%
\bibitem [{\citenamefont {Morita}\ and\ \citenamefont
  {Ogawa}(2010)}]{morita_stability_2010}%
  \BibitemOpen
  \bibfield  {author} {\bibinfo {author} {\bibfnamefont {Y.}~\bibnamefont
  {Morita}}\ and\ \bibinfo {author} {\bibfnamefont {T.}~\bibnamefont {Ogawa}},\
  }\href {\doibase 10.1088/0951-7715/23/6/007} {\bibfield  {journal} {\bibinfo
  {journal} {Nonlinearity}\ }\textbf {\bibinfo {volume} {23}},\ \bibinfo
  {pages} {1387} (\bibinfo {year} {2010})}\BibitemShut {NoStop}%
\bibitem [{\citenamefont {Kuramoto}(2003)}]{kuramoto2003chemical}%
  \BibitemOpen
  \bibfield  {author} {\bibinfo {author} {\bibfnamefont {Y.}~\bibnamefont
  {Kuramoto}},\ }\href@noop {} {\emph {\bibinfo {title} {Chemical oscillations,
  waves, and turbulence}}}\ (\bibinfo  {publisher} {Courier Corporation},\
  \bibinfo {year} {2003})\BibitemShut {NoStop}%
\bibitem [{\citenamefont {Guckenheimer}\ and\ \citenamefont
  {Holmes}(2013)}]{guckenheimer2013nonlinear}%
  \BibitemOpen
  \bibfield  {author} {\bibinfo {author} {\bibfnamefont {J.}~\bibnamefont
  {Guckenheimer}}\ and\ \bibinfo {author} {\bibfnamefont {P.}~\bibnamefont
  {Holmes}},\ }\href@noop {} {\emph {\bibinfo {title} {Nonlinear oscillations,
  dynamical systems, and bifurcations of vector fields}}},\ Vol.~\bibinfo
  {volume} {42}\ (\bibinfo  {publisher} {Springer Science \& Business Media},\
  \bibinfo {year} {2013})\BibitemShut {NoStop}%
\end{thebibliography}
\end{document}